\documentclass[preprint,epsf,aps,floats]{revtex4}
\usepackage{amsmath,amssymb,bm,epsfig,psfrag}
\usepackage{mathtools}

\usepackage{color}

\DeclarePairedDelimiter\abs{\lvert}{\rvert}%
\newcommand\+{\dagger}

\newcommand{\be}{\begin{equation}}
\newcommand{\ee}{\end{equation}}
\newcommand{\ber}{\begin{eqnarray}}
\newcommand{\eer}{\end{eqnarray}}%

\newcommand\bra[1]{\langle #1 |}
\newcommand\ket[1]{|{#1}\rangle}

\def\Dsl{\,\raise.15ex \hbox{/}\mkern-12.8mu D}

\begin{document}

\title{Multiple Exciton Generation in Chiral Carbon Nanotubes: Density Functional Theory Based Computation}
\author{Andrei~Kryjevski,~Deyan~Mihaylov}
\affiliation{Department of Physics,~North Dakota State University,~Fargo, ND~58108,~USA}
\author{Svetlana~Kilina,~Dmitri~Kilin}
\affiliation{Department of Chemistry,~North Dakota State University,~Fargo, ND~58108,~USA}

\begin{abstract}
We use Boltzmann transport equation (BE) to study time evolution of a photo-excited state in a nanoparticle  
including phonon-mediated exciton relaxation and the multiple exciton generation (MEG) processes, such as 
exciton-to-biexciton multiplication and biexciton-to-exciton recombination. 
BE collision integrals are computed using Kadanoff-Baym-Keldysh many-body perturbation 
theory (MBPT) 
based on density functional theory (DFT) simulations, including exciton effects. 
We compute internal quantum efficiency (QE), which is the number of excitons generated from an absorbed
photon in the course of the relaxation. We apply this approach to chiral single-wall carbon nanotubes (SWCNTs), such as (6,2), 
and (6,5).
We predict efficient MEG in the (6,2) and (6,5) SWCNTs
within the solar spectrum range starting at the $2 E_g$ energy threshold and with QE reaching $\sim 1.6$ at
about $3 E_g,$ where $E_g$ is the electronic gap. 
\end{abstract}

\date{\today}

\maketitle

\section{Introduction}
Efficiency of photon-to-electron energy conversion in nanomaterials 
is an important issue that has been actively investigated in recent years. 
For instance, it is envisioned that
efficiency of the nanomaterial-based solar cells can be increased due to 
carrier multiplication, or multiple exciton generation (MEG) process, where 
absorption of a single energetic photon results in the generation of several excitons \cite{10.1063/1.1736034,ISI:000229120900009,AJ2002115}. 
In the course of MEG the excess photon energy is diverted into creating additional charge carriers 
instead of generating atomic vibrations \cite{AJ2002115}. In fact, phonon-mediated electron relaxation 
is a major time-evolution channel competing with the MEG. The conclusion about MEG efficiency in a nanoparticle can only be made 
by simultaneously including MEG, {phonon-mediated carrier relaxation}, and, potentially, other processes, 
such as charge and energy transfer \cite{PhysRevB.88.155304,doi:10.1021/jz4004334}.

In the bulk semiconductor materials MEG in the solar photon energy 
range is known to be inefficient \cite{5144200,5014421,10.1063/1.370658}.
In contrast, in nanomaterials MEG is expected to be enhanced by spatial confinement, which increases electrostatic electron interactions
\cite{doi:10.1146/annurev.physchem.52.1.193,AJ2002115,doi:10.1021/nl0502672,doi:10.1021/nl100177c,doi:10.1021/ar300189j}. 
%
Internal quantum efficiency (QE) is the average number of excitons generated from an absorbed photon. It is a potent measure of MEG efficiency.
QE exceeding 100\% has been measured in recent experiments on, {\it e.g.}, silicon and germanium nanocrystals, and in nanoparticle-based solar cells  \cite{Semonin16122011,doi:10.1021/nl100343j,doi:10.1021/acs.nanolett.5b03161,PbSenanorodQE,GeMEGNature2015,trinh-2012}.

Also, MEG has been observed in single-wall carbon nanotubes (SWCNTs) using transient absorption spectroscopy \cite{doi:10.1021/nl100343j} 
and the photocurrent spectroscopy \cite{Gabor11092009};
$QE=1.3$ at the photon energy $\hbar \omega = 3E^{opt}_g,$ where $E^{opt}_g$ is the optical gap, was found in the (6,5) SWCNT.
Theoretically, MEG in SWCNTs has been investigated using tight-binding approximation with QE up to 
$1.5$ predicted 
{in (17,0) zigzag SWCNT} \cite{PhysRevB.74.121410,PhysRevLett.108.227401}.
{It is understood that in semiconductor nanostructures MEG is dominated by the impact 
ionization process \cite{PhysRevLett.106.207401,PhysRevB.86.165319}. Therefore, MEG QE 
requires calculations of the exciton-to-biexciton decay rate (${\rm R}_{1\to2}$) and of 
the biexciton-to-exciton recombination rate (${\rm R}_{2\to1}$), the direct Auger process, 
and, of course, inclusion of carrier phonon-mediated relaxation. In quasi one-dimensional systems, 
such as SWCNTs, accurate description of these processes requires inclusion of the electron-hole 
bound state effects -- excitons \cite{doi:10.1021/acs.chemrev.5b00012}. 
%
} 
%
%
%

Recently, Density Functional Theory (DFT) combined with the many-body 
perturbation theory (MBPT) techniques has been used to calculate ${\rm R}_{1\to2}$ and ${\rm R}_{2\to1}$ rates, and the photon-to-bi-exciton, 
${\rm R}_2$, and photon-to-exciton, ${\rm R}_1$, rates in two chiral (6,2) and (10,5) SWCNT which have different diameters,
including exciton effects \cite{doi:10.1063/1.4963735}. QE was then estimated as $QE=({\rm R}_1+2 {\rm R}_2)/({\rm R}_1+{\rm R}_2).$ 
However, phonon relaxation was not included. 
Also, ${\rm R}_{1\to2}$ and ${\rm R}_{2\to1}$ for the singlet fission (SF) channel of MEG for these systems have been computed in \cite{SFJCP}. 
The results suggested that efficient MEG in chiral SWCNTs, both in all-singlet and SF channels, 
might be present within the solar spectrum range with ${\rm R}_{1\to2}\sim 10^{14}~s^{-1}$,
while ${\rm R}_{2\to1}/{\rm R}_{1\to2}\leq 10^{-2};$ it was estimated that $QE\simeq 1.2-1.6.$ 
However, MEG strength in these SWCNTs was found to vary strongly with the excitation energy due to highly 
non-uniform density of states. In contrast, the MEG rate calculation done for (6,2) with Cl atoms adsorbed to the surface 
indicated that MEG efficiency in these systems could be enhanced by altering the 
low-energy electronic spectrum via, {\it e.g.}, surface functionalization \cite{SFJCP}. 

Carrier multiplication and recombination without exciton effects and phonon-mediated relaxation was studied using coupled rate equations in 
coupled silicon nanocrystals in \cite{doi:10.1021/ja5057328}.

In this work we develop and apply a Boltzmann transport equation (BE) approach to study MEG in chiral SWCNTs including both exciton 
multiplication and recombination, {\it i.e.}, the Auger processes, and the phonon-mediated exciton relaxation. The Kadanoff-Baym-Keldysh, or NEGF, formalism - a 
generalization of DFT-based MBPT for a non-equilibrium state - allows one to use perturbation theory to calculate collision integrals in 
the transport equation for time evolution of a weakly non-equilibrium photoexcited state \cite{Kadanoff,Keldysh:1964ud,Landau10}. 
Notably, while BE coefficients can be computed to a given order in MBPT, BE itself is non-perturbative.
We calculate QE for the (6,2) and (6,5) SWCNTs.
{This work aims to provide insights into dynamics of photoexcited SWCNTs and its 
dependence on the chirality, diameter and excitation energy.

The paper is organized as follows. Section \ref{sec:method} contains description of the methods
and approximations used in this work. Section III
contains description of the atomistic models used in this work and of the simulation details. 
Section \ref{sec:results} contains discussion of the results obtained. Conclusions and Outlook 
are presented in Section \ref{sec:conclusions}. 
\section{Theoretical Methods and Approximations}
\label{sec:method}
\subsection{Electron Hamiltonian in the KS basis and Approximations}
\label{sec:H}
The the electron field operator $\psi_{\alpha}({\bf x})$ and the annihilation operator of the 
$i^{th}$ Kohn-Sham (KS) state ${\rm a}_{i\alpha}$ are related via
\ber
\psi_{\alpha}({\bf x})=\sum_i\phi_{i\alpha}({\bf x}){\rm a}_{i\alpha},
\label{psi_to_a}
\eer
where $\phi_{i\alpha}({\bf x})$ is the $i^{th}$ KS orbital, and $\alpha$ is the electron spin index \cite{FW,Mahan}. Here we only 
consider spin non-polarzed states with $\phi_{i\uparrow}=\phi_{i\downarrow}\equiv \phi_i;$ also
${\{}{\rm a}_{i\alpha},~{\rm a}_{j\beta}^{\+}{\}}=\delta_{ij}\delta_{\alpha\beta},~{\{}{\rm a}_{i\alpha},~{\rm a}_{j\beta}{\}}=0.$
Then, in 
terms of ${\rm a}_{i\alpha},~{\rm a}_{i\alpha}^{\+},$ the Hamiltonian of electrons in a CNT is (see, {\it e.g.}, \cite{molphysKK,doi:10.1063/1.4963735,SFJCP}.
\ber
{\rm H}=
\sum_{i\alpha}\epsilon_{i} {\rm a}_{i\alpha}^{\+}{\rm a}_{i\alpha}
+{\rm H}_{C}-{\rm H}_{V}+{\rm H}_{e-exciton}.
\label{H}
\eer
where $\epsilon_{i\uparrow}=\epsilon_{i\downarrow}\equiv \epsilon_i$ is the $i^{th}$ KS energy eigenvalue. Typically, in a periodic structure $i={\{}n,{\bf k}{\}},$ where $n$ is the band number, ${\bf k}$ is the lattice wavevector. However, for reasons explained in Section III here KS states are labeled by just integers.
The 
second term is the (microscopic) Coulomb interaction operator 
\ber
{\rm H}_C=\frac12\sum_{ijkl~\alpha,\beta}{\rm V}_{ijkl}{\rm a}^{\dagger}_{i\alpha}{\rm a}^{\dagger}_{j\beta}{\rm a}_{k\beta}{\rm a}_{l\alpha},
~{\rm V}_{ijkl}=\int{\rm d}{\bf x}{\rm d}{\bf y}~\phi^{*}_i({\bf x})\phi^{*}_j({\bf y})\frac{e^2}{|{\bf x}-{\bf y}|}\phi_k({\bf y})
\phi_l({\bf x}).
\label{HC}
\eer
The ${\rm H}_{V}$ term prevents double-counting of electron interactions
\ber
{\rm H}_{V}=\sum_{ij}{\rm a}_{i\alpha}^{\+}\left(\int{\rm d}{\bf x}{\rm d}{\bf y}~\phi^*_i({\bf x}){V_{KS}({\bf x},{\bf y})}\phi_j({\bf y})\right){\rm a}_{j\alpha},
\label{HV}
\eer
where $V_{KS}({\bf x},{\bf y})$ is the KS potential consisting of the Hartree and exchange-correlation terms (see, {\it e.g.}, \cite{RevModPhys.74.601,RevModPhys.80.3}). Photon and electron-photon coupling terms are not directly relevant to this work and, so, are not shown, for brevity.

In this work we will consider singlet excitons only, leaving triplet exciton effects for future work. Before discussing the 
last term in the Hamiltonian (\ref{H}) let us recall that in the Tamm-Dancoff approximation a spin-zero exciton state can be 
represented as 
\cite{PhysRevB.62.4927,PhysRevB.29.5718}
\ber
\ket{{\alpha}}_{0}={\rm B}^{\alpha\dagger}\ket{g.s.}=\sum_{e h}\sum_{\sigma=\uparrow,\downarrow}\frac{1}{\sqrt{2}} {\rm \Psi}^{\alpha}_{eh} a^{\dagger}_{e\sigma} a_{h\sigma} \ket{g.s.},
\label{Psialpha}
\eer 
where the index ranges are $e>HO,~h\leq HO,$ where HO 
is the highest occupied
KS level, $LU=HO+1$ is the lowest unoccupied KS level, and where ${\rm B}^{\alpha\dagger},~{\rm \Psi}^{\alpha}_{eh}$ 
are the singlet {exciton~creation~operator and wavefunction, respectively.}
Then, the last term in the Hamiltonian (\ref{H}) is
\ber
{\rm H}_{e-exciton}&=&
\sum_{e h \alpha}\sum_{\sigma} \frac{1}{\sqrt{2}}\left(\left[\epsilon_{eh}-E^{\alpha}\right]{\rm \Psi}^{\alpha}_{eh}a_{h\sigma}a^{\dagger}_{e\sigma}({\rm B}^{\alpha}+{\rm B}^{\alpha\dagger})+h.c.\right)+\nonumber \\
&+&\sum_{\alpha}E^{\alpha}{\rm B}^{\alpha\dagger}{\rm B}^{\alpha},~\epsilon_{eh}=\epsilon_{e}-\epsilon_{h},
\label{Heexc}
\eer
where $E^{\alpha}$ 
are the singlet {exciton energies.}
The ${\rm H}_{e-exciton}$ term can be seen as the result of re-summation of perturbative 
corrections to the propagating electron-hole state whereby exciton states emerge instead of electron-hole pairs (see, {\it e.g.}, \cite{Berestetskii:1979aa,Beane:2000fx}); it describes coupling of excitons to electrons and holes, 
which allows systematic inclusion of excitons in the perturbative calculations 
\cite{PhysRevLett.92.077402,PhysRevLett.92.257402,PhysRevLett.95.247402,Beane:2000fx}. See \cite{SFJCP} for more details.
To determine
exciton wave functions and energies one solves the Bethe-Salpeter equation (BSE) \cite{PhysRevB.62.4927,PhysRevB.29.5718}.
In the static screening approximation commonly used for semiconductor nanostructures (see, {\it e.g.}, \cite{PhysRevLett.90.127401,PhysRevB.68.085310,PhysRevB.79.245106}) the BSE is \cite{PhysRevB.68.085310}
{
\ber
&&\left(\epsilon_{eh}-E^{\alpha}\right){\rm \Psi}^{\alpha}_{eh}+
\sum_{e^{'}h^{'}} ({\rm K}_{Coul}+{\rm K}_{dir})(e,h;e^{'},h^{'}){\rm \Psi}^{\alpha}_{e^{'},h^{'}}=0,\nonumber \\
&&{\rm K}_{Coul}
=
\sum_{{\bf q}\neq 0}
\frac{8\pi e^2{{\rho}}_{eh}({\bf q}){{\rho}}^{*}_{e^{'}h^{'}}({\bf q})}{V{q}^2}, ~
{\rm K}_{dir}
=
-\frac{1}{V}\sum_{{\bf q}\neq 0}
\frac{4\pi e^2{{\rho}}_{e e^{'}}({\bf q}){{\rho}}^{*}_{h h^{'}}({\bf q})}{{q}^2-\Pi(0,-{\bf q},{\bf q})},
\label{BSEspin0}
\eer}
where
\ber
{{\rho}}_{ji}({\bf p})=\sum_{{\bf k}}\phi_j^{*}({\bf k}-{\bf p})\phi_i({\bf k}),
\label{rhoij}
\eer
is the transitional density, and
\ber
\Pi(\omega,{\bf k},{\bf p})&=&\frac{8 \pi e^2}{V\hbar}\sum_{ij}\rho_{ij}({\bf k})\rho_{ji}({\bf p})\left(\frac{\theta_{-j}\theta_{i}}
{\omega-\omega_{ij}+i\gamma}-\frac{\theta_{j}\theta_{-i}}{\omega-\omega_{ij}-i\gamma}\right),\nonumber \\
\hbar\omega_{ij}&=&\epsilon_{ij},~\sum_{i}\theta_i=\sum_{i > HO},~\sum_{i}\theta_{-i}=\sum_{i \leq HO},
\label{Piwkp}
\eer
is the RPA polarization insertion (see, {\it e.g.}, \cite{FW}).

A major screening approximation used in this work (see Eq. (\ref{BSEspin0}) and Eq. (\ref{VC}))
is that $\Pi(0,{\bf k},{\bf p})\simeq\Pi(0,-{\bf k},{\bf k})\delta_{{\bf k},-{\bf p}}$ corresponding to 
$\Pi(0,{\bf x},{\bf x^{'}})\simeq\Pi(0,{\bf x}-{\bf x^{'}}),$
{\it i.e.},
to the uniform medium approximation. See \cite{doi:10.1063/1.4963735,SFJCP} for more details and discussion of applicability to SWCNTs.



Here, we have used hybrid Heyd-Scuseria-Ernzerhof (HSE06) exchange correlation functional in our DFT simulations \cite{vydrov:074106,heyd:219906} as it has been 
successful in reproducing electronic gaps in various semiconductor nanostructures ({\it e.g.}, 
\cite{MUSCAT2001397,RevModPhys.80.3,doi:10.1021/ct500958p}). (See, however, \cite{PhysRevLett.107.216806}.) Therefore, using the HSE06 functional is to 
substitute for $GW$ corrections to the KS energies -- the first step in the standard three-step 
procedure \cite{PhysRevB.34.5390,PhysRevB.62.4927}. So, 
single-particle energy levels and wave functions are approximated by the KS $\epsilon_i$ and 
$\phi_i({\bf x})$ from the HSE06 DFT output. 
While inclusion of $GW$ corrections would improve accuracy of our calculations, it is unlikely to alter our results and conclusions qualitatively. 

Next, we describe how BE follows from the Keldysh MBPT which is the approach to the description of time evolution of a photoexcited state used in this work. 

\subsection{Boltzmann transport equation from Kadanoff-Baym-Keldysh formalism}
\label{sec:BE}

Description of time evolution of a photoexcited nanoparticle should be comprehensive 
and, thus, include dynamics of electrons, phonons and, also, photons, if one aims to include absorption and recombination. 
For instance, to study MEG one needs to allow carrier multiplication to ``compete" with the phonon-mediated relaxation, 
and, possibly, other processes, such as the energy and charge transfer. 

Boltzmann transport equation (BE) 
is a suitable approach to this problem. The Kadanoff-Baym-Keldysh, or NEGF, formalism - a generalization of MBPT for 
non-equilibrium states - allows one to calculate collision integrals in the transport equation for time evolution of a weakly non-equilibrium
photoexcited state \cite{Kadanoff,Keldysh:1964ud,Landau10}. As noted above, while BE collision integrals can be computed to a given order in MBPT, 
BE itself provides non-perturbative description.

Typically, a simple relaxation time approximation is used for the 
collision integral  \cite{Landau10} (see, {\it e.g.}, \cite{PhysRevLett.108.167402} for a recent application). However, it is 
known that the equation of motion for the Keldysh propagator $G^{-+}$ in the quasi-classical limit reduces to the 
transport equation \cite{Landau10}. {This allows for systematic calculations of the collision integrals using Keldysh MBPT.} 

Let us consider correlation function {$G^{-+}_{\alpha}(t_1,t_2)=\bra{{\cal N}}B^{\dagger}_{\alpha}(t_2)B_{\alpha}(t_1)\ket{{\cal N}},$ 
where $\ket{{\cal N}}$} is some weakly non-equlibrium state, such as a photoexcited state at finite temperature, $B_{\alpha}(t)$ is the 
$\alpha^{th}$ exciton state Heisenberg operator. Note that $G^{-+}_{\alpha}$ is related to the density matrix. The equation of motion for $G^{-+}_{\alpha}(t_1,t_2)$ is
\ber
\left(i\frac{d}{dt_1} - \frac{E^{\alpha}}{\hbar}\right)G^{-+}_{\alpha}(t_1,t_2)=\int{\rm d}{t_3}
\left(\Sigma_{\alpha}^{- -}(t_1,t_3)G_{\alpha}^{-+}(t_3,t_2)+\Sigma_{\alpha}^{- +}(t_1,t_3)G_{\alpha}^{++}(t_3,t_2)\right),
\label{eom}
\eer
where $\Sigma_{\alpha} ^{-+},~\Sigma_{\alpha}^{--},~G_{\alpha}^{++}$ are the Keldysh self-energies and correlators, respectively.
BE obtains when the slow {$t=(t_1+t_2)/2$} and fast (intrinsic) {$t_0=t_1-t_2$} times are introduced (see paragraph 95 of \cite{Landau10} for details). 
It describes slow time-evolution of a weakly non-equilibrium state.
For instance, the $t_1 \to t_2$ components of $G_{\alpha}^{-+}$ are the exciton occupation numbers: 
$n_{\alpha }(t)=G_{\alpha}^{-+}(t,t).$ It is the set of $n_{\alpha }(t)$ that describes the non-equilibrium state $\ket{{\cal N}}.$  
Then BE for the (slow) time evolution of a photo-excited state is
\ber
\dot{n}_{\alpha}=i\Sigma_{\alpha}^{-+}\left(n;\omega _{\alpha}\right)\left(1+n_{\alpha }\right)-i\Sigma_{\alpha}^{+-}\left(n;\omega _{\alpha }\right)n_{\alpha },~
\omega_{\alpha }=\frac{E^{\alpha}}{\hbar},
\label{BEq}
\eer
where $\Sigma_{\alpha} ^{-+},~\Sigma_{\alpha}^{+-}$ are the leading Keldysh exciton self-energies, which depend on the slowly varying occupation numbers. 
Different contributions to the self-energies that correspond to different processes will be discussed below. 
Note the r.h.s. of Eq. \ref{BEq} has the expected ``gain'' - ``loss'' term structure. 
The approach is applicable if $\dot{n}_{\alpha}(t)\ll \omega_{\alpha},$ which is the quasi-classicality condition in this case. 
Solving the system of equations (\ref{BEq}) with the initial condition 
$n_{\alpha}(t=0) \neq 0$ for $\hbar\omega_{\alpha}=\hbar\omega,$ where the excitation energy $\hbar\omega$ corresponds to, {\it e.g.}, an 
absorption peak, will yield description of relaxation in a photo-excited nanoparticle.

As mentioned above, here we aim to describe dynamics of a photoexcited SWCNT including 
electron-hole bound state (exciton) effects and taking into account 
(I)~phonon-mediated relaxation, {\it i.e.}, the non-adiabatic processes;
(II)~exciton-to-bi-exciton decay and bi-exciton recombination, {\it i.e.}, inverse and direct Auger processes.
Collision integrals for these two processes will be described in the next two sub-sections.

Then, in order to calculate internal QE one \\
1.~populates exciton with energy $E;$ the initial state is $n_{i}(0)=n_{in}\delta_{i\alpha},~E_{\alpha}=E$, \\
2.~solves BE including a.~phonon emission and absorption terms, b.~exciton-to-bi-exciton decay and recombination, \\
3.~adds up the occupation numbers of excitons generated after the occupancies 
have plateaued as $t \to \infty$. (Recall that recombination occurring on a much longer time scale is not included here.) After the initial state averaging, 
\ber
QE(E)=\frac{\sum_{\alpha} n_{\alpha}}{n_{in}}.
\label{QE}
\eer

\subsection{Electron-Phonon Interaction}
\label{sec:enu}

In order to include phonon terms and the electron-phonon coupling Hamiltonian (\ref{H}) is augmented with
\ber
{\rm H}_{N}=\sum_{I=1}^{{\rm N}_{ion}}\frac{P_I^2}{2 M_I}+V_{NN},~V_{NN}\simeq\frac12\sum_{I\neq J}\frac{{\rm Z}_I{\rm Z}_J e^2}{|{\bf R}_I-{\bf R}_J|},
\label{HRP_ion} 
\eer
where ${\bf P}_I,~{\bf R}_I,~M_I,~Z_I$ are the $I^{th}$ ion momentum, position, mass, and effective charge -- number of valence electrons, respectively; ${\bf P}_I,~{\bf R}_I$ are the ion momentum and position operators, respectively, $I=1,..,N_{ion}.$
One sets ${\bf R}_I={\bf R}_I^0+{\bf r}_I,$ where the equilibrium ion positions ${\bf R}_I^0$ corresponding to the energy minimum have been found by, {\it e.g.}, the DFT geometry relaxation procedure, and ${\bf r}_I$ are small oscillations about equilibrium. This leads to the following phonon and electron-phonon terms in the Hamiltonian (see, {\it e.g.}, \cite{PhysRevB.91.085305,doi:10.1021/bk-2015-1196.ch010})
\ber 
\delta{\rm H}_{e-ph}=\sum_{\nu=1}^{3{\rm N}_{ion}-6}\hbar\omega_{\nu}\left({\rm c}_{\nu}^{\dagger}{\rm c}_{\nu}+\frac12\right)+
\sum_{ij\nu\sigma}{\rm g}^{\nu}_{ij}{\rm a}_{i\sigma}^{\+}{\rm a}_{j\sigma}({\rm c}_{\nu}^{\dagger}+{\rm c}_{\nu}),
\label{Henu}
\eer
where $c_{\nu}$ is the phonon annihilation operator, and the electron-phonon couplings are \cite{doi:10.1021/bk-2015-1196.ch010}
\ber
{{\rm g}}^{\nu}_{ij}=\sum_{I=1}^{{\rm N}_i}~\frac{4 \pi i {\rm Z}_I e^2}{V}\sqrt{\frac{\hbar}{2 \omega_{\nu}M_I}}\sum_{{\bf p}}
\frac{\rho^{*}_{ji}({\bf p})\left({\bf p}\cdot{\bf U}_I^{\nu}\right){\rm e}^{-i{\bf p}\cdot{\bf R}_I}}{p^2-\Pi(0,-{\bf p},{\bf p})}.
\label{gijnu}
\eer
This is similar to the frozen photon approximation \cite{vanCamp,PhysRevB.85.235422}. Our approach is applicable to the systems where 
atoms undergo small oscillations about their equilibrium positions; the electronic states are approximated using equilibrium atomic 
positions ${\bf R}_I^0.$ Normal frequencies, $\omega_{\nu}$, and mode decompositions, ${\bf U}^{\nu}_I$, are calculated using 
DFT software, such as VASP. The pseudopotential ${\rm v}({\bf x},{\bf R}_I)$ felt by the valence electrons is approximated here by the 
Coulomb interaction, which is also (approximately) screened \cite{FW}.

The ${\cal O}({r}_I^2)$ contribution to $\delta{\rm H}_{e-ph}$ (see, {\it e.g.}, \cite{ANDP:ANDP201000100}) is neglected since it does not contribute to the 
exciton-phonon Keldysh couplings at the leading one-pnonon level. More generally, the ${\cal O}({r}_I^2)$ term's contribution is suppressed due to 
screening of ${\rm v}({\bf x},{\bf R}_I)$ \cite{FW,AGD}. 

\begin{figure}[!t]
\center
\begin{tabular}{cc}
\raisebox{0.21875\totalheight}{\includegraphics*[width=0.45\textwidth] 
{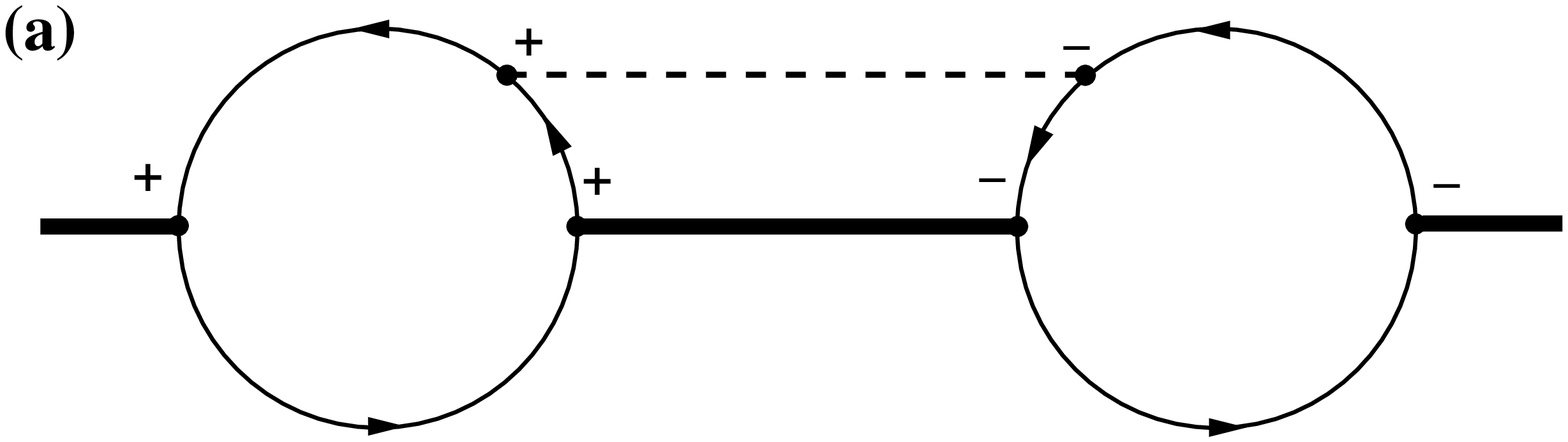}} &
\raisebox{0.35\totalheight}{\includegraphics*[width=0.535\textwidth] 
{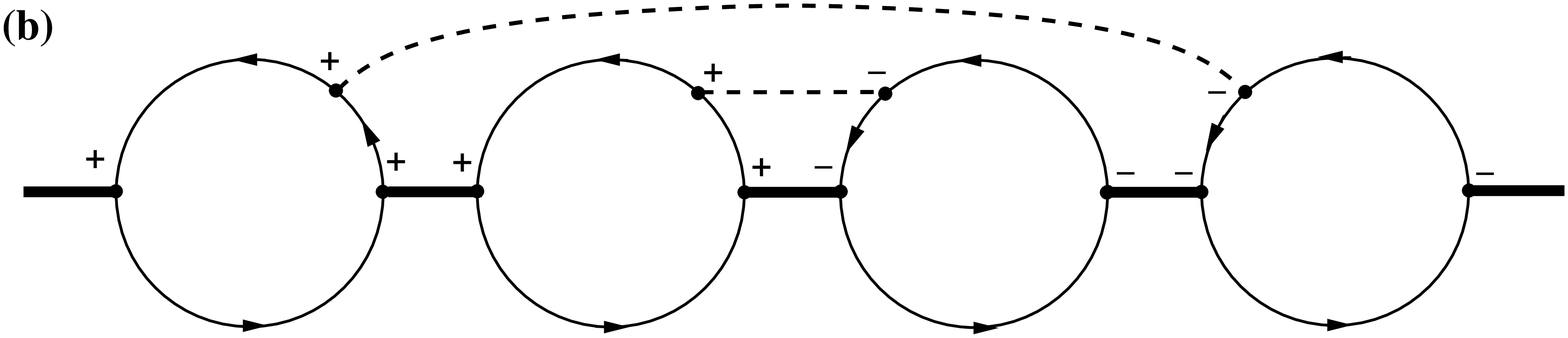}}\\
\end{tabular}
\vspace{-5.25ex}
\caption{Typical Feynman diagrams for the $\Sigma^{+-}$ describing exciton-phonon coupling. Shown in (a) and (b) are the one- and two-phonon processes, respectively. Thin solid lines stand for the KS state propagators, thick solid lines depict excitons, dashed lines -- phonons. 
The -+ contributions obtain when + and - are interchanged.
}
\label{fig:Gnu12}
\end{figure}
So, in order to describe phonon-mediated relaxation in a nanoparticle one includes electron-phonon interactions described 
above in Eqs. (\ref{Henu},\ref{gijnu}). The corresponding Feynman diagrams are shown in Fig. \ref{fig:Gnu12}. In SWCNTs sub-gaps 
in the electronic energy spectrum are sizable enough for the two-phonon processes (Fig. \ref{fig:Gnu12},~(b)) to be relevant.
With the one- and two-phonon processes included the BE is
\ber 
&& \dot{n}_{\alpha} =\sum_{\alpha \mu} (G_{1})_{\alpha \beta }^{\mu } \left(n_{\beta } n_{\mu }-n_{\alpha }
   \left(n_{\beta }+n_{\mu }+1\right)\right) \delta \left(\omega _{\alpha
   }-\omega _{\beta }-\omega _{\mu }\right) + \nonumber \\
&& + \sum_{\alpha \mu \nu}({G}_{2})_{\alpha \beta }^{\mu \nu } \left(n_{\mu } n_{\nu }
   \left(n_{\beta }-n_{\alpha }\right)-n_{\alpha } \left(n_{\beta
   }+1\right) \left(n_{\mu }+n_{\nu }\right)-n_{\alpha } \left(n_{\beta
   }+1\right)\right) \times \nonumber \\
&&\times\delta \left(\omega _{\alpha }-\omega _{\beta }-\omega
   _{\mu }-\omega _{\nu }\right) - \{\alpha \leftrightarrow\beta\},~\text{where}
\label{BEph}
\eer
\ber
n_{\nu} \simeq \left({\rm exp}\left[\frac{\hbar\omega_{\nu}}{k_B T}\right]-1\right)^{-1}, 
\label{nnu}
\eer
and
\ber
 (G_{1})_{\alpha \beta }^{\mu } \simeq \frac{2\pi}{\hbar^2}\left(\abs*{\sum_{ijk}\theta _{-i} \theta _j \theta _k g_{{jk}}^{\mu } \left(\Psi
   _{{ji}}^{\alpha }\right){}^* \Psi _{{ki}}^{\beta }}^2+\abs*{\sum_{ijk}\theta _i \theta _{-j} \theta _{-k} g_{{jk}}^{\mu } 
\Psi_{{ij}}^{\alpha } \left(\Psi_{{ik}}^{\beta }\right){}^*}^2\right),
\label{1phonon}
\eer
where $(G_{1})_{\alpha \beta }^{\mu }$ are the dominant terms of the effective exciton-phonon couplings, T is the temperature.  
Also,
\ber
&& (G_{2})_{\alpha \beta }^{\mu \nu} \simeq \frac{2\pi}{\hbar^4}\abs*{\sum_{pkqjli}\sum_{\sigma}\frac{1}{\omega_{\alpha }+\omega_{\mu }-\omega_{\sigma }+i \delta}
\left(c_{pp}+c_{hh}+c_{ph}+c_{hp}\right)}^2,\nonumber \\
&& c_{pp}=g_{{pk}}^{\mu } g_{{qj}}^{\nu }\frac{\theta _{-k} \theta _l \theta _{-p} \left(\Psi _{{lk}}^{\alpha
   }\right){}^* \Psi _{{lp}}^{\sigma } \left(\omega _{{lp}}-\omega
   _{\sigma }\right)}{\omega_{{lp}}-\omega_{\alpha } -\omega_{\mu }-i \delta}\frac{\theta _i \theta _{-j} \theta _{-q} \left(\Psi _{\text{ij}}^{\sigma
   }\right){}^* \left(\omega_{{ij}}-\omega_{\sigma }\right) \Psi_{{iq}}^{\beta }}{\omega_{{ij}}-\omega _{\alpha } -\omega _{\mu }-i \delta},\nonumber \\
&& c_{hh}=g_{{pj}}^{\mu } g_{{ql}}^{\nu }\frac{\theta _{-i} \theta _j \theta _p \Psi _{{ji}}^{\alpha }
   \left(\Psi _{pi }^{\sigma }\right){}^* \left(\omega _{pi }-\omega_{\sigma }\right)}
{\omega _{\alpha } +\omega _{\mu }-\omega_{pi }+i \delta}\frac{\theta _{-k} \theta _l \theta _q \Psi _{\text{lk}}^{\sigma }
   \left(\omega _{{lk}}-\omega _{\sigma }\right) 
\left(\Psi_{{qk}}^{\beta }\right){}^*}{\omega _{\alpha } +\omega_{\mu }-\omega _{{lk}}+i \delta},\nonumber \\
&& c_{hp}=g_{{jq}}^{\nu} g_{{pk}}^{\mu}\frac{\theta _k \theta _{-l} \theta _p \Psi _{{kl}}^{\alpha }
   \left(\Psi _{{pl}}^{\sigma }\right){}^* \left(\omega_{{pl}}-\omega _{\sigma }\right)}{\omega _{\alpha } +\omega_{\mu }-\omega _{{pl}}+i \delta}
\frac{\theta_i \theta_{-j} \theta_{-q} \Psi _{{ij}}^{\sigma }
   \left(\omega _{{ij}}-\omega _{\sigma }\right) \left(\Psi_{{iq}}^{\beta }\right){}^*}{\omega_{{ij}}-\omega _{\alpha } -\omega _{\mu }-i \delta},\nonumber \\
&& c_{ph}=g_{{kp}}^{\mu} g_{{qj}}^{\nu}\frac{\theta _{-k} \theta _l \theta _{-p} \Psi _{{lk}}^{\alpha }
   \left(\Psi _{{lp}}^{\sigma }\right){}^* \left(\omega_{{lp}}-\omega_{\sigma }\right)}{\omega_{{lp}}-\omega _{\alpha}-\omega_{\mu}-i \delta}
\frac{\theta _{-i} \theta _j \theta _q \Psi _{{ji}}^{\sigma }
   \left(\omega _{{ji}}-\omega _{\sigma }\right) 
\left(\psi_{{qi}}^{\beta }\right){}^*}
{\omega _{\alpha} -\omega_{{ji}}+\omega _{\mu}+i \delta},
\label{2phonon}
\eer
where $\delta$ is the small width parameter set here to the temperature scale $25~meV.$
Both phonon emission and absorption from the thermal bath by either the electron or the hole within the exciton are included in Eq. (\ref{BEph}).
The two-phonon processes where one phonon is emitted and the other one is absorbed 
are not included in Eq. (\ref{BEph}), for technical simplicity. 
Also, in this perturbative approach phonon occupation numbers are approximated by their equilibrium values (see Eq. (\ref{nnu})), which is another technical simplification.

{In the above and in the following expressions only the terms leading in 
the ratio of the typical exciton binding energy to the gap $(\epsilon_{binding}/E_g) < 1$ are shown, for brevity. But full expressions have been included in the actual calculations.}

\subsection{MEG terms}
\label{sec:meg}

Transport equations that describe exciton-to-biexciton decay and biexciton-to-exciton recombination are 
{
\ber 
& \frac{\text{dn}_{\gamma} }{\text{dt}} =\sum_{\alpha \beta}R_{\alpha \beta}^{\gamma } \left(n_{\alpha } n_{\beta }-n_{\gamma }
   \left(n_{\alpha }+n_{\beta }+1\right)\right) \delta \left(\omega _{\alpha
   }+\omega _{\beta }-\omega _{\gamma }\right),\nonumber \\
&n_{\alpha } \frac{\text{dn}_{\beta } }{\text{dt}}+n_{\beta }\frac{\text{dn}_{\alpha }
   }{\text{dt}}=-\sum_{\gamma}R_{\alpha \beta}^{\gamma } \left(n_{\alpha } n_{\beta }-n_{\gamma }
   \left(n_{\alpha }+n_{\beta }+1\right)\right) \delta \left(\omega _{\alpha
   }+\omega _{\beta }-\omega _{\gamma }\right),
\label{BeqMEG}
\eer
}
where $R_{\alpha \beta}^{\gamma }$ are the rates from the MEG contributions to $\Sigma_{\alpha}^{-+},~\Sigma_{\alpha}^{+-}$ shown in Fig. \ref{fig:R1to2}. In (\ref{BeqMEG}) the first equation describes the I.I. process where exciton $\gamma$ decays into excitons $\alpha$ and $\beta;$ the second -- inverse recombination process (direct Auger).
\begin{figure}[!t]
\center
\raisebox{0.385\totalheight}{\includegraphics*[width=0.975\textwidth] 
{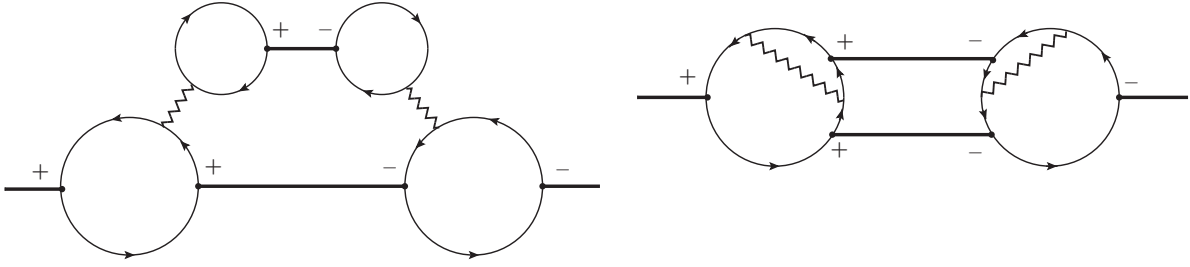}}
\vspace{-.75in}
\caption{Exciton self-energy Feynman diagrams relevant for the exciton$\to$bi-exciton and bi-exciton$\to$exciton processes. Thin solid lines stand for the KS state propagators, thick solid lines depict excitons, zigzag lined -- screened Coulomb potential. The diagrams on the left and the right correspond to the exchange and direct channels, respectively. Not shown for brevity are the similar diagrams with all the Fermion arrows reversed and with + and - interchanged.
}
\vspace{-.1in}
\label{fig:R1to2}
\end{figure}
The leading order rate expressions are ($\alpha, \beta, \gamma$ subscripts are omitted for brevity) \cite{doi:10.1063/1.4963735}
\ber
R_{\alpha \beta}^{\gamma }&=&\left(R^p+R^h+{\tilde R}^p+{\tilde R}^h\right)\delta(\omega_{\gamma}-\omega_{\alpha} -\omega_{\beta}),\nonumber \\
 R^p&=&2\frac{2 \pi}{\hbar^2}
\abs*{\sum_{ijkln}W_{{jlnk}}
\theta_l \theta_{-n} 
(\Psi_{ln }^{\beta }){}^*
\theta_i \theta_{-j} \theta_{-k}
\Psi_{{ij}}^{\gamma} 
\left(\Psi_{{ik}}^{\alpha }\right){}^*
}^2,\nonumber \\
R^h&=&2\frac{2 \pi}{\hbar^2}
\abs*{\sum_{ijkln}W_{{jlnk}} 
\theta_{-l} \theta_n 
\Psi_{{nl}}^{\beta }
\theta _{-i} \theta_j \theta _k
   (\Psi_{{ji}}^{\gamma }){}^* \Psi_{{ki}}^{\alpha}
}^2.
\label{R1to2all}
\eer
The expressions for ${\tilde R}^h$ and ${\tilde R}^p$ are the same as the ones for $R^h,~R^p$ with $W_{jlnk}$ replaced by $W_{jlkn}$ and divided by 2.
In the above
\ber
W_{jlnk}&=&\sum_{{\bf q}\neq 0}\frac{4 \pi e^2}{V}\frac{{{\rho}}_{kj}^{*}({\bf q}){{\rho}}_{ln}({\bf q})}
{\left(q^2-\Pi(0,-{\bf q},{\bf q})\right)}
\label{VC}
\eer
is the (approximate) screened Coulomb matrix element.

Strictly speaking, here we are working to the second order in the 
screened Coulomb interaction. This refers to the electron$\to$trion, hole$\to$trion sub-processes in Fig. \ref{fig:R1to2}, {\it i.e.}, the trion is created in the course of a single Coulomb interaction. However, as discussed above, the electron-hole interactions that form the excitons are included to all orders. Also, in this work a biexciton state is approximated by a pair of non-interacting excitons.

  
{\section{Computational Details}}
\label{sec:compdetail}
The optimized SWCNT geometries, KS orbitals and energy eigenvalues 
have been obtained using the {\it{ab initio}} total energy and
molecular dynamics program VASP (Vienna ab initio simulation program) 
with the hybrid Heyd-Scuseria-Ernzerhof (HSE06) exchange correlation functional \cite{vydrov:074106,heyd:219906}
using the 
projector augmented-wave (PAW) pseudopotentials \cite{PhysRevB.50.17953,PhysRevB.59.1758}.

Applying conjugated gradient method for atomic position relaxation the 
structures were relaxed until residual forces on the ions 
were no greater than $0.05~eV/\AA.$ 
The momentum cutoff defined by
\ber
\frac{\hbar^2{k}^2}{2 m}\leq {\cal E}_{max},
\label{Ecutoff}
\eer
where 
$m$ is the electron mass, was chosen at ${\cal E}_{max}=400~eV.$
The energy cutoffs determined by the number of KS orbitals included in the simulations
were chosen so that 
$\epsilon_{i_{max}}-\epsilon_{HO}\simeq\epsilon_{LU}-\epsilon_{i_{min}}\geq 3~eV,$ 
where $i_{max},~i_{min}$ are the highest and the lowest KS labels included in simulations.
\begin{figure}[!t]
\center
\vspace{-1.5cm}
\includegraphics*[width=0.995\textwidth]{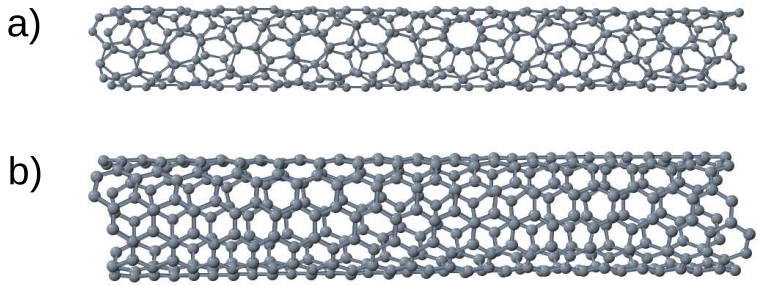}
\vspace{-17.4cm}
\caption{Atomistic models of chiral SWCNTs. Shown in a) is SWCNT (6,2). 
Three unit cells have been included in the simulations. In b) is SWCNT (6,5). Only one unit cell is included due to computational cost restrictions. 
}
\vspace{-0.45ex}
\label{fig:optimized_structures}
\end{figure}
SWCNT atomistic models 
were placed in the finite volume simulation boxes with periodic boundary conditions. In the axial direction the box size was chosen to accommodate 
an integer number of unit cells. In the other two directions the SWCNTs 
have been kept separated by about $1~nm$ of vacuum surface-to-surface which 
excluded spurious interactions between their periodic images.
\begin{figure}[!t]
\vspace{-5.25ex}
\center
\begin{tabular}{cc}
{\bf a)} & {\bf b)} \\
\raisebox{0.02875\totalheight}{\includegraphics*[width=0.485\textwidth] 
{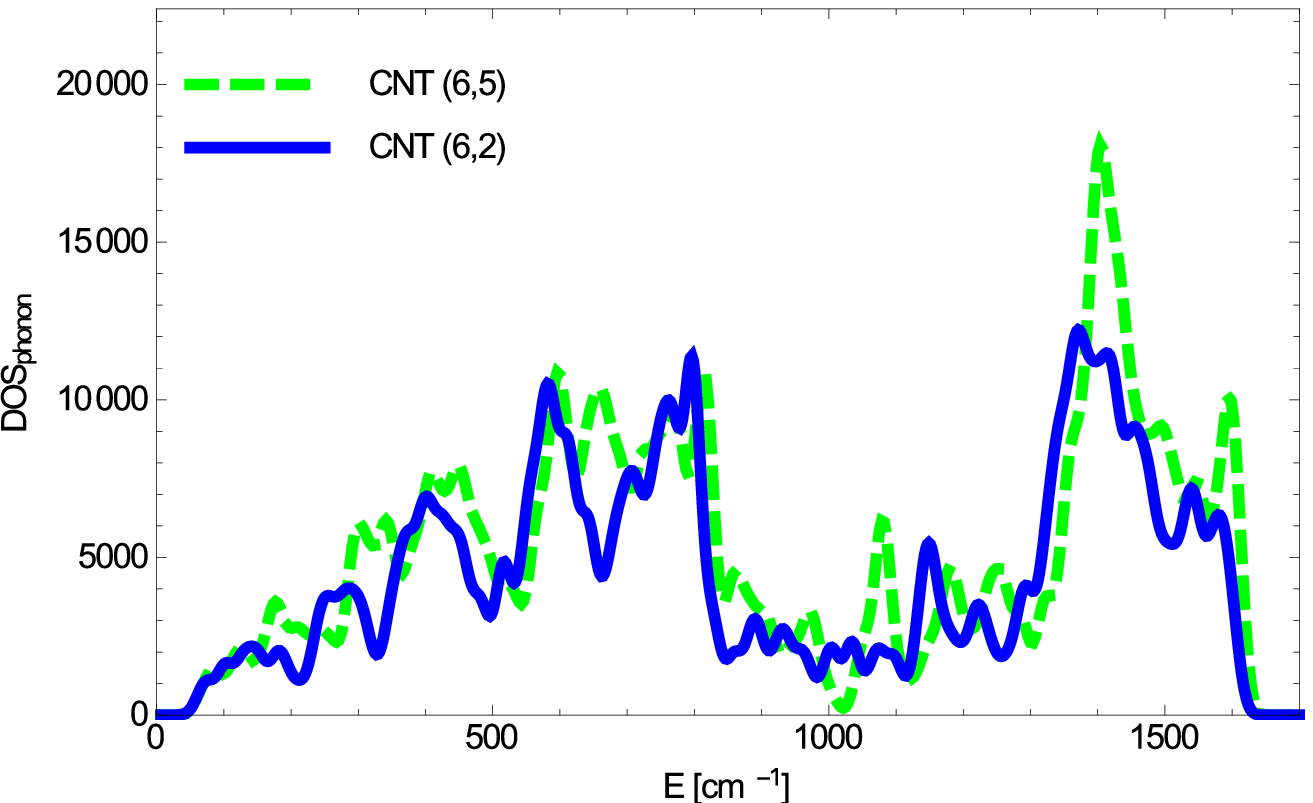}} &
\raisebox{0.02175\totalheight}{\includegraphics*[width=0.485\textwidth] 
{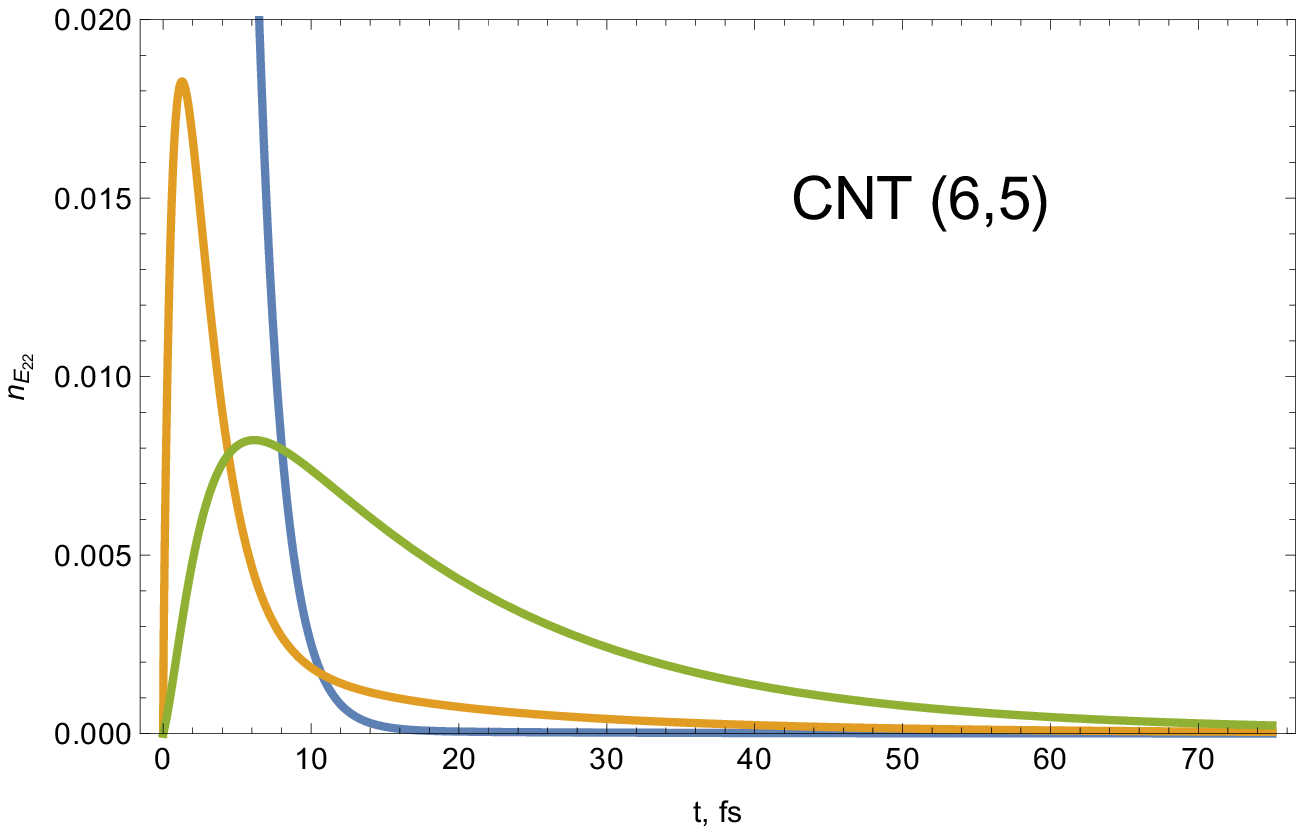}}\\
{\bf c)} & {\bf d)}\\ 
\raisebox{1.6375\totalheight}{\includegraphics*[width=0.485\textwidth] 
{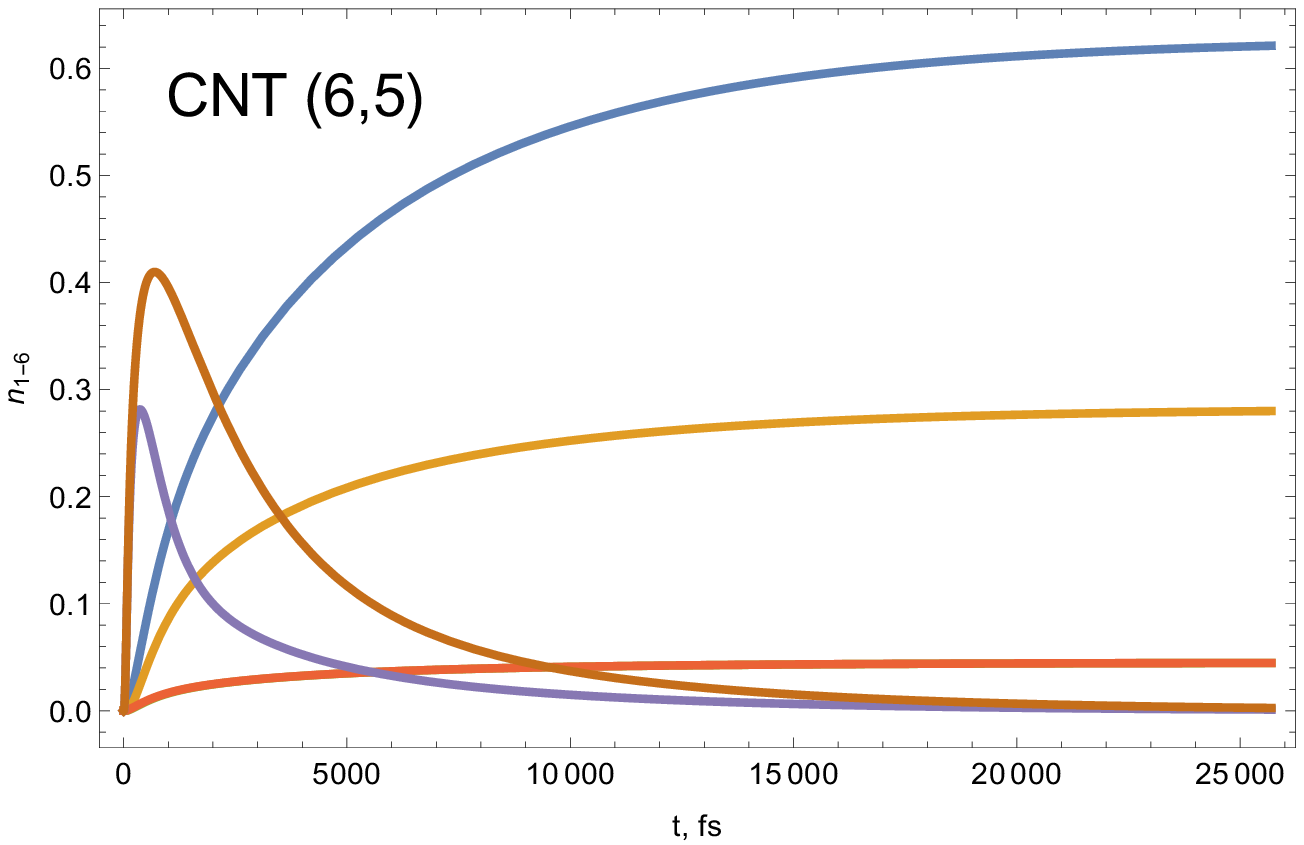}}&
\raisebox{1.6375\totalheight}{\includegraphics*[width=0.485\textwidth] 
{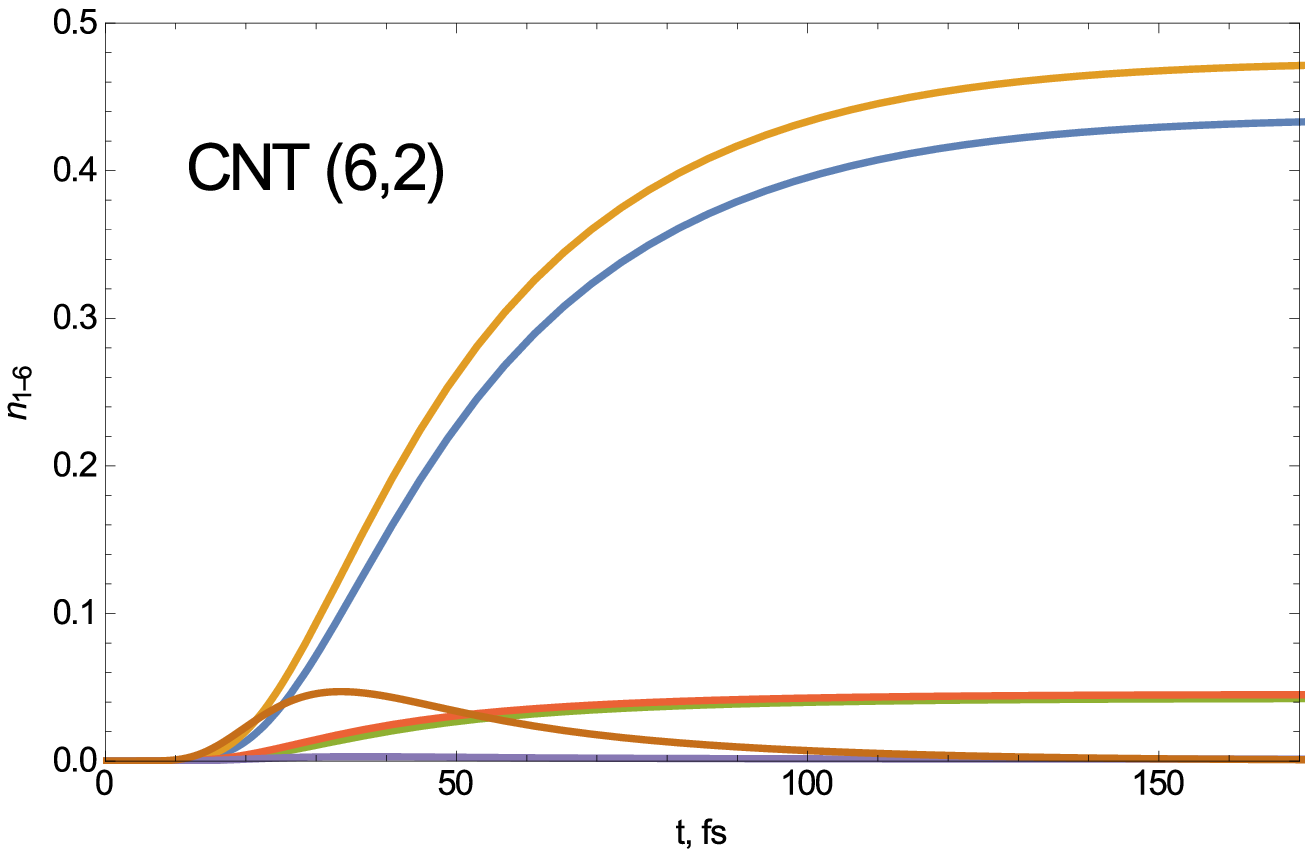}} \\
\end{tabular}
\vspace{-3.25in}
\caption{Phonon DOSs for the two CNTs are shown in a). Shown in b) are the occupation numbers of the exciton states corresponding to the $E_{22}$ peak energy in CNT (6,5). In c) and d) are the few low-energy exciton occupancies after excitation at the $E_{22}$ peak energy in (6,5) and (6,2), respectively. }
\label{fig:62_vs_65}
\end{figure}
As discussed in \cite{doi:10.1063/1.4963735}, we have found reasonably small (about 10\%) variation in the single 
particle energies over the Brillouin zone when three unit cells were included in the DFT simulations \cite{doi:10.1063/1.4963735}.
Therefore, in (6,2) simulations have been done at the $\Gamma$ point including three unit cells.
In our approximation lattice momenta of the KS states, which are suppressed by the reduced Brillouin zone size, have been neglected. 
Due to high computational cost, for (6,5) SWCNT only one unit cell was included. However, we checked that this 
size-reduced model reproduced the absorption spectrum features with the accuracy similar to other SWCNTs \cite{SFJCP}. 
As explained before in \cite{doi:10.1063/1.4963735}, the rationale for including multiple unit cells instead of the Brillouin zone sampling 
in the DFT simulations is that surfaces of these SWCNTs are to be functionalized.
Inclusion of several unit cells keeps the concentration of surface dopants reasonably low. 
\begin{figure}[!t]
\vspace{-8.25ex}
\center
\begin{tabular}{cc}
{\bf a)} & {\bf b)} \\
\raisebox{0.02875\totalheight}{\includegraphics*[width=0.485\textwidth] 
{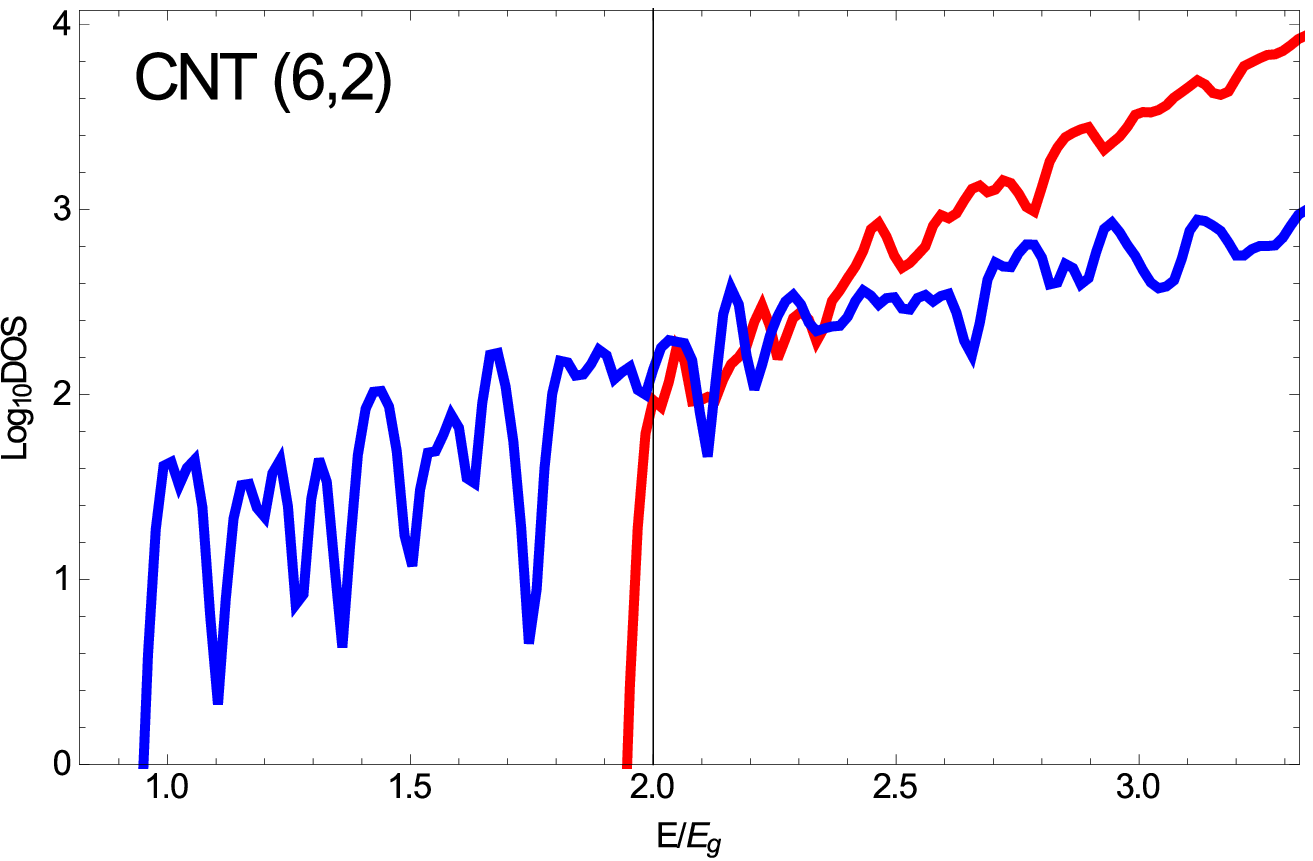}} &
\raisebox{0.02175\totalheight}{\includegraphics*[width=0.485\textwidth] 
{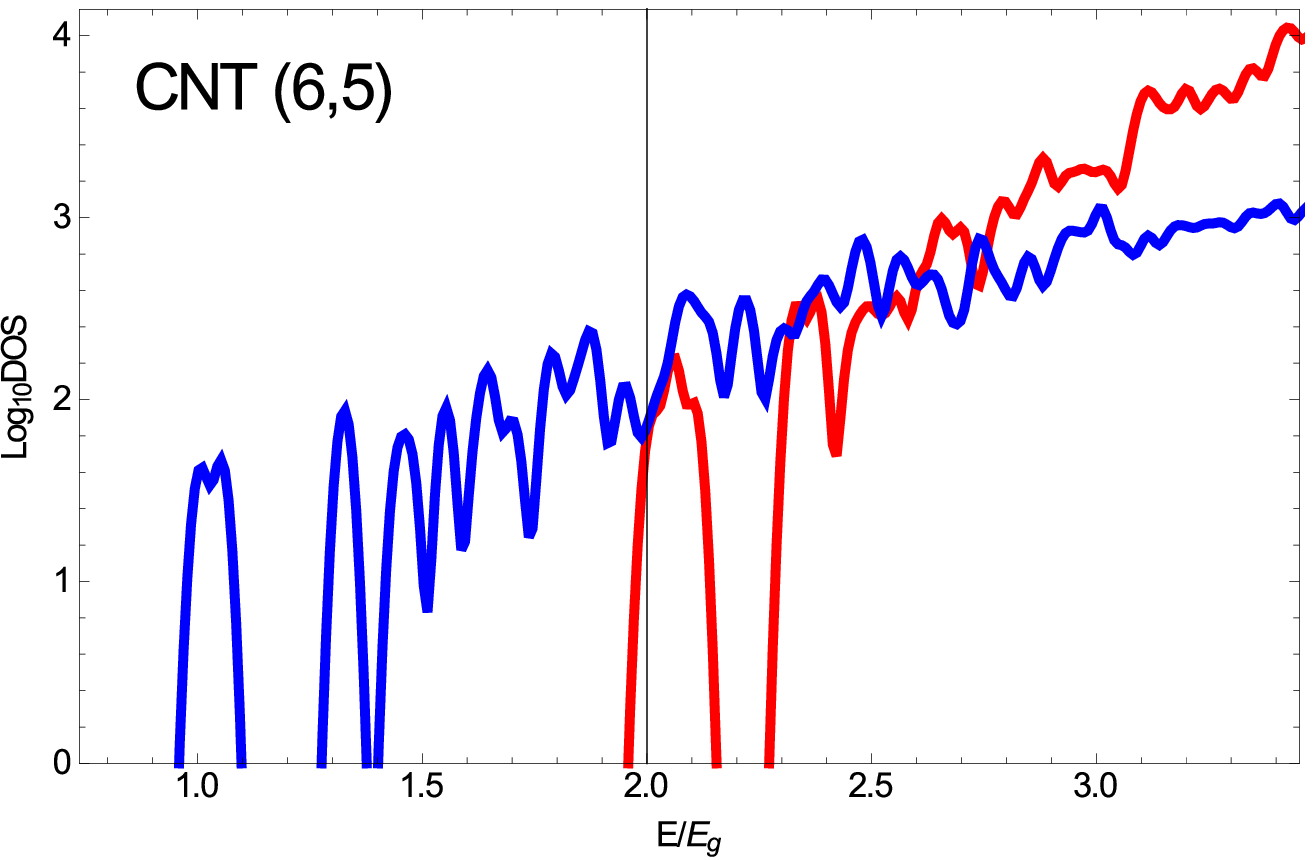}}\\
{\bf c)} & {\bf d)}\\ 
\raisebox{1.6375\totalheight}{\includegraphics*[width=0.485\textwidth] 
{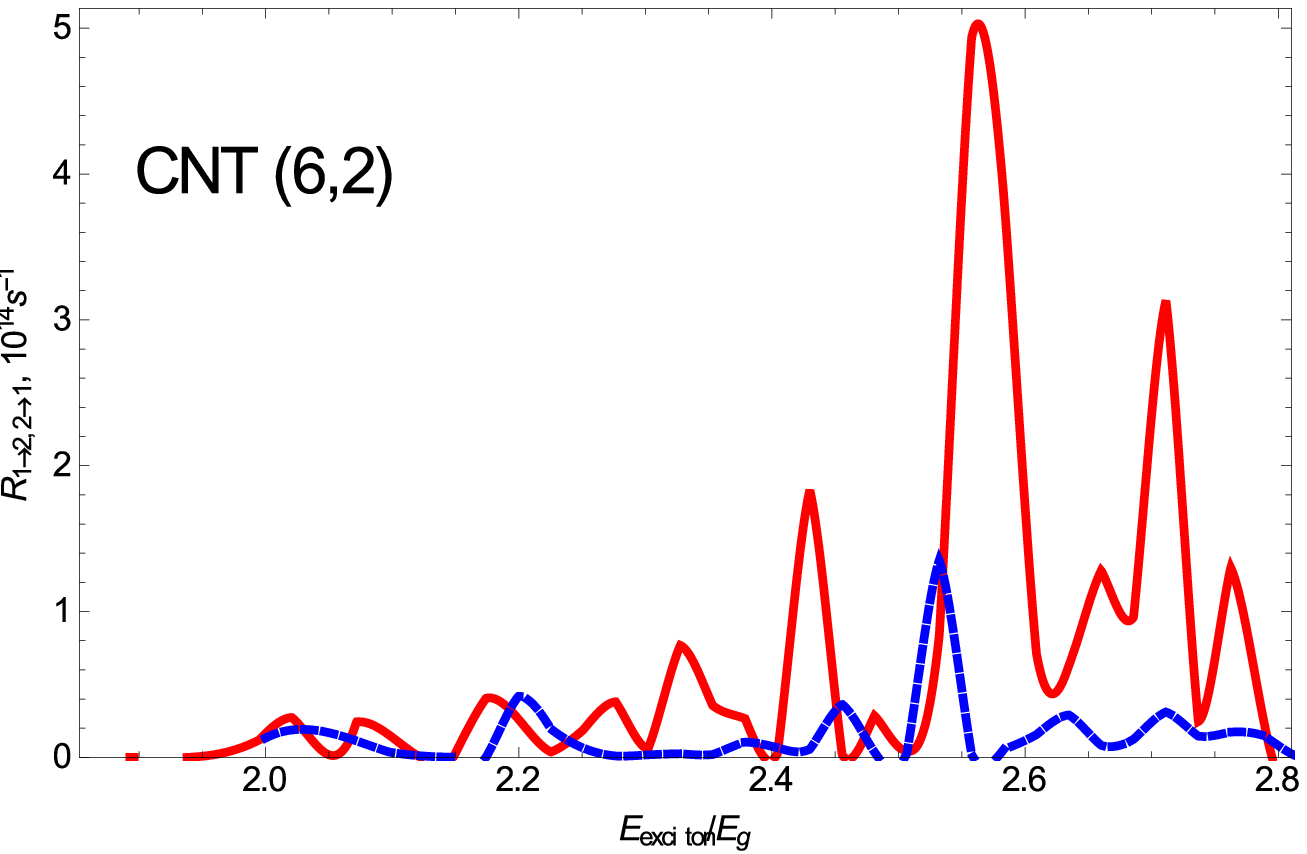}}&
\raisebox{1.61875\totalheight}{\includegraphics*[width=0.485\textwidth] 
{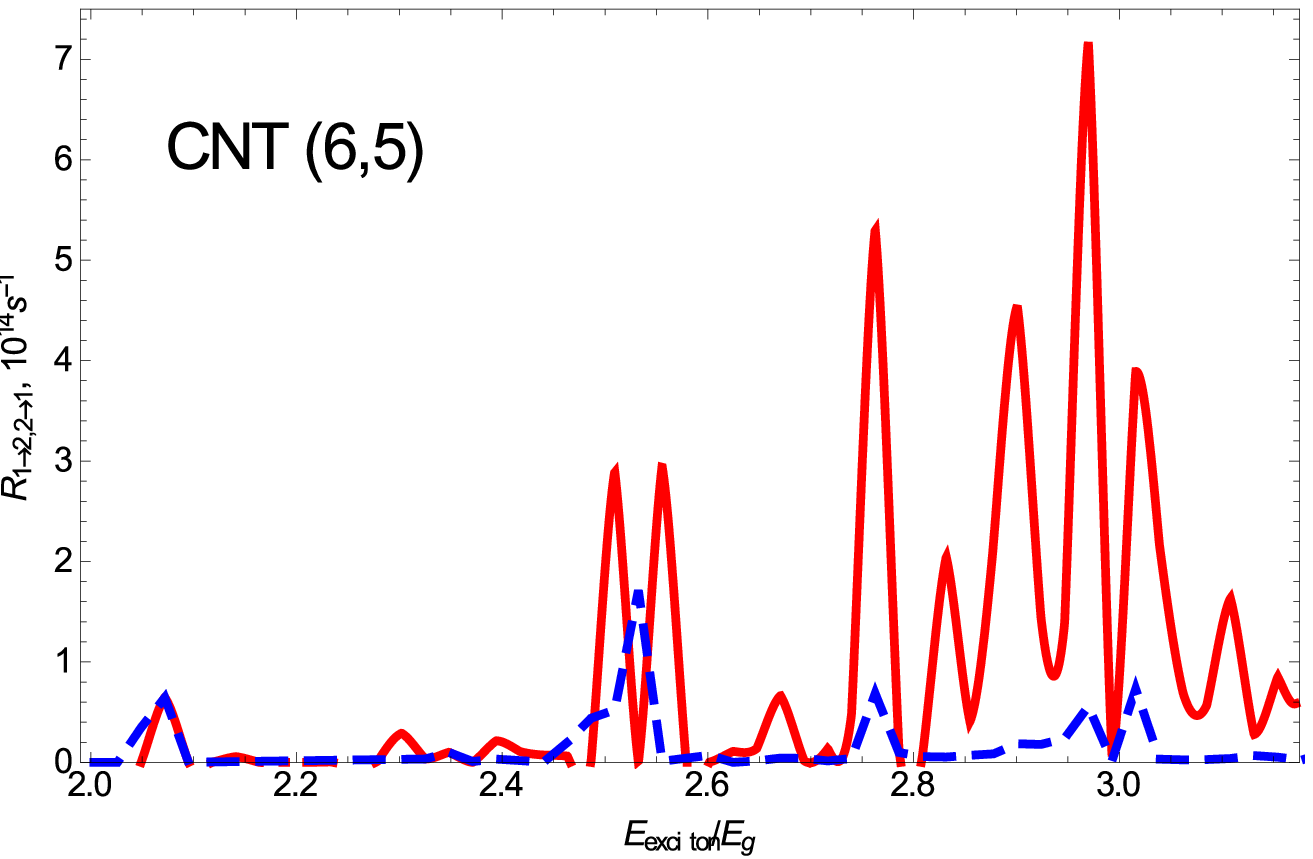}} \\
\end{tabular}
\vspace{-3.295in}
\caption{Exciton and bi-exciton  DOSs for the CNT (6,2) ($E_g = 0.98~eV$) and CNT (6,5) ($E_g = 1.09~eV$) are shown in a) and b), respectively; in c) and d) are 
the exciton-to-biexciton, bi-exciton-to-exciton rates 
$R_{1\to 2},~R_{2\to 1}$ in CNT (6,2) and (6,5), respectively. {\it Note that recombination rate magnitude -- the dashed curves in c) and d) -- has been multiplied by 100. } 
}
\label{fig:DOSR12_62_vs_65}
\end{figure}
The atomistic models of the optimized nanotubes are shown in Fig. (\ref{fig:optimized_structures}).
%
%

In this work all the DFT simulations have been done in a vacuum which is to approximate properties of these SWCNTs 
dispersed in a non-polar solvent. 

\section{Results and Discussion}
\label{sec:results}

First, we have simulated relaxation of an energetic exciton state corresponding to the absorption peak $E_{22}$ including electron-phonon interactions only. 
The results are shown in Fig. (\ref{fig:62_vs_65}). Phonon spectra for CNTs (6,2) and (6,5) computed using VASP are in Fig. (\ref{fig:62_vs_65}),~a). 
Shown in Fig. (\ref{fig:62_vs_65}),~b) are the occupancies of the $E_{22}$ peak exciton states in CNT (6,5). As expected, the initial excitation is cascading 
down in energy. Intermediate states are being excited and subsequently decay in the course of the relaxation.  
An exponential fit to the predicted $n_{E_{22}}(t)$ curve has yielded decay constant $\tau_{22}=16.7~fs.$ The available experimental results are for 
(6,5) in the aqueous solution where $\tau_{22}=120~fs$ was reported \cite{doi:10.1021/nl2038503}. For SWCNTs of unspecified chiralities immersed in 
polyethylene glycol and in polymethylmethacrylate $\tau_{22}\simeq 40~fs$ was reported \cite{PhysRevLett.94.207401,SWCNTtau}. So, while direct comparison 
is not possible at this time, it is likely that our prediction for $\tau_{22}$ is underestimated. This is as expected and is in line with other predictions 
of our approach which tends to overestimate strength of couplings. Overall, this confirms applicability of our method for semi-quantitative description of 
photoexcited chiral SWCNT dynamics. 

Shown in Fig. (\ref{fig:62_vs_65}),~c) and d) are the few low-energy exciton occupancies resulting from the excitation at the $E_{22}$ peak energy for (6,5) and (6,2), respectively.
The excitation goes through several transient states before forming a terminal steady state where only few low-energy levels are occupied. 
In (6,2) the relaxation time from the $E_{22}$ excitation is about 100 fs 
(Fig. (\ref{fig:62_vs_65}), d)), 
while in 
(6,5) the $E_{22}$ relaxation time is predicted to be two orders of magnitude longer (Fig. (\ref{fig:62_vs_65}), c)). 
The results 
suggest that relaxation rates in CNTs strongly depend on the diameter and chirality.

Next, we augmented the system of equations for $n_{\alpha}(t)$ with the MEG terms. Now as the excitation is cascading down the energy levels by emitting and absorbing 
phonons, it can, also, undergo an exciton$\to$bi-exciton decay. Conversely, the bi-exciton state can recombine into a single high-energy exciton. For completeness, the 
exciton and bi-exciton spectra, as well as the exciton$\to$bi-exciton rates from \cite{doi:10.1063/1.4963735} and \cite{SFJCP} are shown in Fig. \ref{fig:DOSR12_62_vs_65}. 
These results suggest that MEG strength in SWCNTs is likely to be determined by an intricate interplay between the Auger processes and phonon-mediated relaxation. For 
instance, just above the $2 E_g$ threshold there are energy ranges where bi-exciton DOS is lower the DOS of single excitons (Fig. \ref{fig:DOSR12_62_vs_65}), a) and b)). Counterintuitively enough, the faster phonon-mediated relaxation will carry the excitation through these energy ranges the smaller fraction of bi-excitons will be able 
to recombine, thus enhancing MEG. Also, we note that for efficient MEG it is crucial to have non-zero $R_{1 \to 2}$ just above the $2 E_g$ threshold. 

Using the QE procedure outlined in Section \ref{sec:BE} we computed QE for the CNTs (6,2) and (6,5) as a function of the excitation energy. 
Our results are shown in Fig. \ref{fig:QE}. We predict efficient solar range MEG in both of these systems which starts at the threshold and reaches 
$\sim 1.6$ at about $3 E_g.$

As already mentioned above, MEG in CNT (6,5) has been studied experimentally in \cite{doi:10.1021/nl100343j}.
QE was measured for two excitation energies:
for {{$E=2.5 E^{opt}_g~QE=1.1$}} was reported, while our prediction is {{$QE \simeq 1.2$}};
for {{$E=3 E^{opt}_g~QE=1.3$}} {\it vs.} {{$QE \simeq 1.6,$}} which is our prediction. The ways to improve accuracy of our method are discussed in Section \ref{sec:conclusions}.
\begin{figure}[!t]
\vspace{-5.25ex}
\center
\raisebox{1.375\totalheight}{\includegraphics*[width=0.625\textwidth] 
{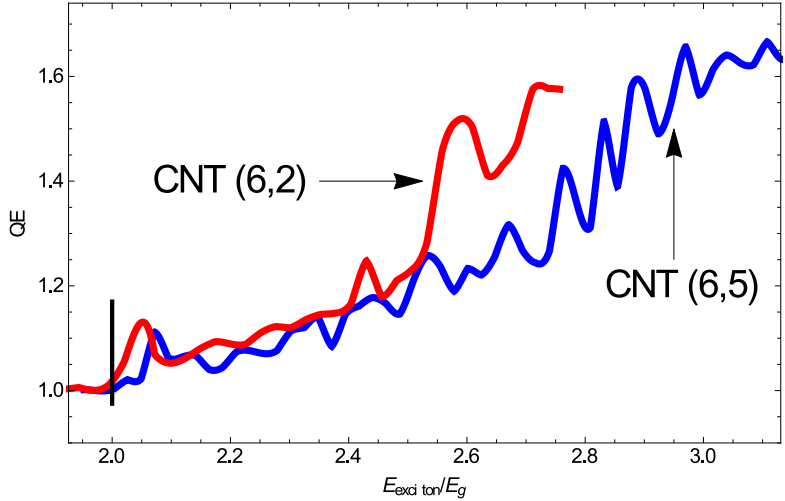}}
\vspace{-3.75in}
\caption{The QEs for the CNT (6,2) and (6,5) {\it vs.} $E/E_g,$ where $E_g = 0.98~eV$ for (6,2) and for (6,5) $E_g = 1.09~eV.$ Vertical dash indicates $2 E^{BSE}_g$ threshold.
}
\label{fig:QE}
\end{figure}
\section{Conclusions and Outlook}
\label{sec:conclusions}
We have used Kadanoff-Baym-Keldysh, or NEGF, technique of MBPT to develop a comprehensive DFT-based description of
time evolution of a photoexcited SWCNT. The approach is based on the Boltzmann transport equation for time evolution of 
a weakly non-equilibrium photoexcited state in the quasi-classical limit with collision integrals calculated 
from the DFT-based Keldysh MBPT.  
We have been working to the second order in the RPA-screened Coulomb interaction 
and including electron-hole bound state effects for which we had to solve BSE. 

This method has been used to study MEG in the chiral SWCNTs, 
using (6,2) and (6,5) as examples. In particular, we calculated predictions for the internal QE as a function of the excitation energy.
Our calculations suggest that chiral SWCNTs may have efficient MEG within the solar spectrum range (see Fig. \ref{fig:QE}).

In the pristine SWCNTs the MEG rates vary strongly with the excitation energy. In contrast, using $Cl$-decorated (6,2) SWCNT as an example
it has been found that surface functionalization significantly alters low-energy spectrum in a SWCNT \cite{SFJCP}. Also, in the doped case, 
the MEG rate is not only greater in magnitude, but also is a much smoother function of the excitation energy.  
QE calculations for the doped SWCNTs and for different chiral SWCNTs are in progress.

As described above, several simplifying approximations had to be utilized in order to be able to calculate properties of these SWCNTs. 
Previously, we checked have that our predictions for the absorption spectra were in a reasonable agreement with the experimental data 
with the error less then 13\% for E$_{11}$ and E$_{22}$ peak energies for the (6,2), (6,5) and (10,5) nanotubes. As discussed in Section 
\ref{sec:results}, in this work we compared our predictions for the phonon-mediated relaxation and for $QE(E)$ to the available experimental 
results for CNT (6,5). The comparison indicated that our current method is valid for SWCNTs at the semi-quantitative level.  
However, accuracy can be improved in several ways. For instance, static interaction approximation $\Pi(\omega,{\bf k},{\bf p})={\Pi}(0,{\bf k},{\bf p})$ 
is reasonably accurate when employed in the BSE. But in the impact ionization process the typical energy exchange is of order of the gap and, so, 
screening should be treated as dynamical. This is likely to enhanced screening (see Eq. (\ref{Piwkp}) for $\omega\simeq E_g$) which could help alleviate 
the overbinding issue. Implementation of this is technically challenging and is left to future work.
A straightforward improvement is to include $GW$ single particle energy corrections, which then can be easily incorporated in the 
rate expressions. {It is likely to blue-shift the rate curves 
by a fraction of eV without significant changes to the overall shape.} Another natural but technically challenging step is to use full RPA 
interaction ${\rm W}(\omega,{\bf k},{\bf p})$ rather than ${\rm W}(0,-{\bf k},{\bf k}).$ 

However, none of these corrections are likely to change the main results and conclusions of this work,
while drastically increasing computational cost.

\section{Acknowledgments}
Authors acknowledge financial support from the NSF grant CHE-1413614. 
S.K. acknowledges partial support of the Alfred P. Sloan Research Fellowship BR2014-073 and NDSU Advance FORWARD
program sponsored by NSF HRD-0811239 and EPS-0814442.
The authors acknowledge the use of computational resources at
the Center for Computationally Assisted Science and
Technology (CCAST) at North Dakota State University and
the National Energy Research Scientific Computing Center
(NERSC) allocation award 86678, supported by the Office of
Science of the DOE under contract No. DE-AC02-05CH11231.



\bibliography{dft}

\begin{thebibliography}{63}
\expandafter\ifx\csname natexlab\endcsname\relax\def\natexlab#1{#1}\fi
\expandafter\ifx\csname bibnamefont\endcsname\relax
  \def\bibnamefont#1{#1}\fi
\expandafter\ifx\csname bibfnamefont\endcsname\relax
  \def\bibfnamefont#1{#1}\fi
\expandafter\ifx\csname citenamefont\endcsname\relax
  \def\citenamefont#1{#1}\fi
\expandafter\ifx\csname url\endcsname\relax
  \def\url#1{\texttt{#1}}\fi
\expandafter\ifx\csname urlprefix\endcsname\relax\def\urlprefix{URL }\fi
\providecommand{\bibinfo}[2]{#2}
\providecommand{\eprint}[2][]{\url{#2}}

\bibitem[{\citenamefont{Shockley and Queisser}(1961)}]{10.1063/1.1736034}
\bibinfo{author}{\bibfnamefont{W.}~\bibnamefont{Shockley}} \bibnamefont{and}
  \bibinfo{author}{\bibfnamefont{H.}~\bibnamefont{Queisser}},
  \bibinfo{journal}{J. Appl. Phys.} \textbf{\bibinfo{volume}{32}},
  \bibinfo{pages}{510} (\bibinfo{year}{1961}).

\bibitem[{\citenamefont{Ellingson
  et~al.}(2005{\natexlab{a}})\citenamefont{Ellingson, Beard, Johnson, Yu,
  Micic, Nozik, Shabaev, and Efros}}]{ISI:000229120900009}
\bibinfo{author}{\bibfnamefont{R.~J.} \bibnamefont{Ellingson}},
  \bibinfo{author}{\bibfnamefont{M.~C.} \bibnamefont{Beard}},
  \bibinfo{author}{\bibfnamefont{J.~C.} \bibnamefont{Johnson}},
  \bibinfo{author}{\bibfnamefont{P.~R.} \bibnamefont{Yu}},
  \bibinfo{author}{\bibfnamefont{O.~I.} \bibnamefont{Micic}},
  \bibinfo{author}{\bibfnamefont{A.~J.} \bibnamefont{Nozik}},
  \bibinfo{author}{\bibfnamefont{A.}~\bibnamefont{Shabaev}}, \bibnamefont{and}
  \bibinfo{author}{\bibfnamefont{A.~L.} \bibnamefont{Efros}},
  \bibinfo{journal}{Nano Letters} \textbf{\bibinfo{volume}{5}},
  \bibinfo{pages}{865} (\bibinfo{year}{2005}{\natexlab{a}}).

\bibitem[{\citenamefont{Nozik}(2002)}]{AJ2002115}
\bibinfo{author}{\bibfnamefont{A.~J.} \bibnamefont{Nozik}},
  \bibinfo{journal}{Physica E: Low-dimensional Systems and Nanostructures}
  \textbf{\bibinfo{volume}{14}}, \bibinfo{pages}{115 } (\bibinfo{year}{2002}).

\bibitem[{\citenamefont{de~Boer et~al.}(2013)\citenamefont{de~Boer, de~Jong,
  Timmerman, Gregorkiewicz, Zhang, Buma, Poddubny, Prokofiev, and
  Yassievich}}]{PhysRevB.88.155304}
\bibinfo{author}{\bibfnamefont{W.~D. A.~M.} \bibnamefont{de~Boer}},
  \bibinfo{author}{\bibfnamefont{E.~M. L.~D.} \bibnamefont{de~Jong}},
  \bibinfo{author}{\bibfnamefont{D.}~\bibnamefont{Timmerman}},
  \bibinfo{author}{\bibfnamefont{T.}~\bibnamefont{Gregorkiewicz}},
  \bibinfo{author}{\bibfnamefont{H.}~\bibnamefont{Zhang}},
  \bibinfo{author}{\bibfnamefont{W.~J.} \bibnamefont{Buma}},
  \bibinfo{author}{\bibfnamefont{A.~N.} \bibnamefont{Poddubny}},
  \bibinfo{author}{\bibfnamefont{A.~A.} \bibnamefont{Prokofiev}},
  \bibnamefont{and} \bibinfo{author}{\bibfnamefont{I.~N.}
  \bibnamefont{Yassievich}}, \bibinfo{journal}{Phys. Rev. B}
  \textbf{\bibinfo{volume}{88}}, \bibinfo{pages}{155304}
  (\bibinfo{year}{2013}).

\bibitem[{\citenamefont{Stewart et~al.}(2013)\citenamefont{Stewart, Padilha,
  Bae, Koh, Pietryga, and Klimov}}]{doi:10.1021/jz4004334}
\bibinfo{author}{\bibfnamefont{J.}~\bibnamefont{Stewart}},
  \bibinfo{author}{\bibfnamefont{L.}~\bibnamefont{Padilha}},
  \bibinfo{author}{\bibfnamefont{W.}~\bibnamefont{Bae}},
  \bibinfo{author}{\bibfnamefont{W.}~\bibnamefont{Koh}},
  \bibinfo{author}{\bibfnamefont{J.}~\bibnamefont{Pietryga}}, \bibnamefont{and}
  \bibinfo{author}{\bibfnamefont{V.}~\bibnamefont{Klimov}},
  \bibinfo{journal}{The Journal of Physical Chemistry Letters}
  \textbf{\bibinfo{volume}{4}}, \bibinfo{pages}{2061} (\bibinfo{year}{2013}).

\bibitem[{\citenamefont{Bude and Hess}(1992)}]{5144200}
\bibinfo{author}{\bibfnamefont{J.}~\bibnamefont{Bude}} \bibnamefont{and}
  \bibinfo{author}{\bibfnamefont{K.}~\bibnamefont{Hess}}, \bibinfo{journal}{J.
  Appl. Phys.} \textbf{\bibinfo{volume}{72}}, \bibinfo{pages}{3554 }
  (\bibinfo{year}{1992}).

\bibitem[{\citenamefont{Jung et~al.}(1996)\citenamefont{Jung, Taniguchi, and
  Hamaguchi}}]{5014421}
\bibinfo{author}{\bibfnamefont{H.~K.} \bibnamefont{Jung}},
  \bibinfo{author}{\bibfnamefont{K.}~\bibnamefont{Taniguchi}},
  \bibnamefont{and}
  \bibinfo{author}{\bibfnamefont{C.}~\bibnamefont{Hamaguchi}},
  \bibinfo{journal}{J. Appl. Phys.} \textbf{\bibinfo{volume}{79}},
  \bibinfo{pages}{2473} (\bibinfo{year}{1996}).

\bibitem[{\citenamefont{Harrison et~al.}(1999)\citenamefont{Harrison, Abram,
  and Brand}}]{10.1063/1.370658}
\bibinfo{author}{\bibfnamefont{D.}~\bibnamefont{Harrison}},
  \bibinfo{author}{\bibfnamefont{R.~A.} \bibnamefont{Abram}}, \bibnamefont{and}
  \bibinfo{author}{\bibfnamefont{S.}~\bibnamefont{Brand}}, \bibinfo{journal}{J.
  Appl. Phys.} \textbf{\bibinfo{volume}{85}}, \bibinfo{pages}{8186}
  (\bibinfo{year}{1999}).

\bibitem[{\citenamefont{Nozik}(2001)}]{doi:10.1146/annurev.physchem.52.1.193}
\bibinfo{author}{\bibfnamefont{A.}~\bibnamefont{Nozik}},
  \bibinfo{journal}{Annual Review of Physical Chemistry}
  \textbf{\bibinfo{volume}{52}}, \bibinfo{pages}{193} (\bibinfo{year}{2001}).

\bibitem[{\citenamefont{Ellingson
  et~al.}(2005{\natexlab{b}})\citenamefont{Ellingson, Beard, Johnson, Yu,
  Micic, Nozik, Shabaev, and Efros}}]{doi:10.1021/nl0502672}
\bibinfo{author}{\bibfnamefont{R.}~\bibnamefont{Ellingson}},
  \bibinfo{author}{\bibfnamefont{M.}~\bibnamefont{Beard}},
  \bibinfo{author}{\bibfnamefont{J.}~\bibnamefont{Johnson}},
  \bibinfo{author}{\bibfnamefont{P.}~\bibnamefont{Yu}},
  \bibinfo{author}{\bibfnamefont{O.}~\bibnamefont{Micic}},
  \bibinfo{author}{\bibfnamefont{A.}~\bibnamefont{Nozik}},
  \bibinfo{author}{\bibfnamefont{A.}~\bibnamefont{Shabaev}}, \bibnamefont{and}
  \bibinfo{author}{\bibfnamefont{A.}~\bibnamefont{Efros}},
  \bibinfo{journal}{Nano Letters} \textbf{\bibinfo{volume}{5}},
  \bibinfo{pages}{865} (\bibinfo{year}{2005}{\natexlab{b}}).

\bibitem[{\citenamefont{McGuire et~al.}(2010)\citenamefont{McGuire, Sykora,
  Joo, Pietryga, and Klimov}}]{doi:10.1021/nl100177c}
\bibinfo{author}{\bibfnamefont{J.}~\bibnamefont{McGuire}},
  \bibinfo{author}{\bibfnamefont{M.}~\bibnamefont{Sykora}},
  \bibinfo{author}{\bibfnamefont{J.}~\bibnamefont{Joo}},
  \bibinfo{author}{\bibfnamefont{J.}~\bibnamefont{Pietryga}}, \bibnamefont{and}
  \bibinfo{author}{\bibfnamefont{V.}~\bibnamefont{Klimov}},
  \bibinfo{journal}{Nano Letters} \textbf{\bibinfo{volume}{10}},
  \bibinfo{pages}{2049} (\bibinfo{year}{2010}).

\bibitem[{\citenamefont{Gabor}(2013)}]{doi:10.1021/ar300189j}
\bibinfo{author}{\bibfnamefont{N.~M.} \bibnamefont{Gabor}},
  \bibinfo{journal}{Accounts of Chemical Research}
  \textbf{\bibinfo{volume}{46}}, \bibinfo{pages}{1348} (\bibinfo{year}{2013}).

\bibitem[{\citenamefont{Semonin et~al.}(2011)\citenamefont{Semonin, Luther,
  Choi, Chen, Gao, Nozik, and Beard}}]{Semonin16122011}
\bibinfo{author}{\bibfnamefont{O.}~\bibnamefont{Semonin}},
  \bibinfo{author}{\bibfnamefont{J.}~\bibnamefont{Luther}},
  \bibinfo{author}{\bibfnamefont{S.}~\bibnamefont{Choi}},
  \bibinfo{author}{\bibfnamefont{H.-Y.} \bibnamefont{Chen}},
  \bibinfo{author}{\bibfnamefont{J.}~\bibnamefont{Gao}},
  \bibinfo{author}{\bibfnamefont{A.~J.} \bibnamefont{Nozik}}, \bibnamefont{and}
  \bibinfo{author}{\bibfnamefont{M.~C.} \bibnamefont{Beard}},
  \bibinfo{journal}{Science} \textbf{\bibinfo{volume}{334}},
  \bibinfo{pages}{1530} (\bibinfo{year}{2011}).

\bibitem[{\citenamefont{Wang et~al.}(2010)\citenamefont{Wang, Khafizov, Tu,
  Zheng, and Krauss}}]{doi:10.1021/nl100343j}
\bibinfo{author}{\bibfnamefont{S.}~\bibnamefont{Wang}},
  \bibinfo{author}{\bibfnamefont{M.}~\bibnamefont{Khafizov}},
  \bibinfo{author}{\bibfnamefont{X.}~\bibnamefont{Tu}},
  \bibinfo{author}{\bibfnamefont{M.}~\bibnamefont{Zheng}}, \bibnamefont{and}
  \bibinfo{author}{\bibfnamefont{T.}~\bibnamefont{Krauss}},
  \bibinfo{journal}{Nano Letters} \textbf{\bibinfo{volume}{10}},
  \bibinfo{pages}{2381} (\bibinfo{year}{2010}).

\bibitem[{\citenamefont{Bohm et~al.}(2015{\natexlab{a}})\citenamefont{Bohm,
  Jellicoe, Tabachnyk, Davis, Wisnivesky-Rocca-Rivarola, Ducati, Ehrler,
  Bakulin, and Greenham}}]{doi:10.1021/acs.nanolett.5b03161}
\bibinfo{author}{\bibfnamefont{M.~L.} \bibnamefont{Bohm}},
  \bibinfo{author}{\bibfnamefont{T.~C.} \bibnamefont{Jellicoe}},
  \bibinfo{author}{\bibfnamefont{M.}~\bibnamefont{Tabachnyk}},
  \bibinfo{author}{\bibfnamefont{N.~J. L.~K.} \bibnamefont{Davis}},
  \bibinfo{author}{\bibfnamefont{F.}~\bibnamefont{Wisnivesky-Rocca-Rivarola}},
  \bibinfo{author}{\bibfnamefont{C.}~\bibnamefont{Ducati}},
  \bibinfo{author}{\bibfnamefont{B.}~\bibnamefont{Ehrler}},
  \bibinfo{author}{\bibfnamefont{A.~A.} \bibnamefont{Bakulin}},
  \bibnamefont{and} \bibinfo{author}{\bibfnamefont{N.~C.}
  \bibnamefont{Greenham}}, \bibinfo{journal}{Nano Letters}
  \textbf{\bibinfo{volume}{15}}, \bibinfo{pages}{7987}
  (\bibinfo{year}{2015}{\natexlab{a}}).

\bibitem[{\citenamefont{Bohm et~al.}(2015{\natexlab{b}})\citenamefont{Bohm,
  Tabachnyk, Wisnivesky-Rocca-Rivarola, Jellicoe, Ducati, Ehrler, and
  Greenham}}]{PbSenanorodQE}
\bibinfo{author}{\bibfnamefont{M.~L.} \bibnamefont{Bohm}},
  \bibinfo{author}{\bibfnamefont{M.}~\bibnamefont{Tabachnyk}},
  \bibinfo{author}{\bibfnamefont{F.}~\bibnamefont{Wisnivesky-Rocca-Rivarola}},
  \bibinfo{author}{\bibfnamefont{T.~C.} \bibnamefont{Jellicoe}},
  \bibinfo{author}{\bibfnamefont{C.}~\bibnamefont{Ducati}},
  \bibinfo{author}{\bibfnamefont{B.}~\bibnamefont{Ehrler}}, \bibnamefont{and}
  \bibinfo{author}{\bibfnamefont{N.~C.} \bibnamefont{Greenham}},
  \bibinfo{journal}{Nature Communications} \textbf{\bibinfo{volume}{6}},
  \bibinfo{pages}{8259} (\bibinfo{year}{2015}{\natexlab{b}}).

\bibitem[{\citenamefont{Saeed et~al.}(2015)\citenamefont{Saeed, de~Weerd,
  Spoor, Houtepen, Siebbeles, and Gregorkiewicz}}]{GeMEGNature2015}
\bibinfo{author}{\bibfnamefont{S.}~\bibnamefont{Saeed}},
  \bibinfo{author}{\bibfnamefont{P.}~\bibnamefont{de~Weerd},
  \bibfnamefont{C.~Stallinga}},
  \bibinfo{author}{\bibfnamefont{F.}~\bibnamefont{Spoor}},
  \bibinfo{author}{\bibfnamefont{A.}~\bibnamefont{Houtepen}},
  \bibinfo{author}{\bibfnamefont{L.}~\bibnamefont{Siebbeles}},
  \bibnamefont{and}
  \bibinfo{author}{\bibfnamefont{T.}~\bibnamefont{Gregorkiewicz}},
  \bibinfo{journal}{Light: Science and Applications}
  \textbf{\bibinfo{volume}{4}}, \bibinfo{pages}{672} (\bibinfo{year}{2015}).

\bibitem[{\citenamefont{Trinh et~al.}(2012)\citenamefont{Trinh, Limpens,
  de~Boer, Schins, Siebbeles, and Gregorkiewicz}}]{trinh-2012}
\bibinfo{author}{\bibfnamefont{M.~T.} \bibnamefont{Trinh}},
  \bibinfo{author}{\bibfnamefont{R.}~\bibnamefont{Limpens}},
  \bibinfo{author}{\bibfnamefont{W.}~\bibnamefont{de~Boer}},
  \bibinfo{author}{\bibfnamefont{J.}~\bibnamefont{Schins}},
  \bibinfo{author}{\bibfnamefont{L.}~\bibnamefont{Siebbeles}},
  \bibnamefont{and}
  \bibinfo{author}{\bibfnamefont{T.}~\bibnamefont{Gregorkiewicz}},
  \bibinfo{journal}{Nature Photonics} \textbf{\bibinfo{volume}{6}},
  \bibinfo{pages}{316} (\bibinfo{year}{2012}).

\bibitem[{\citenamefont{Gabor et~al.}(2009)\citenamefont{Gabor, Zhong, Bosnick,
  Park, and McEuen}}]{Gabor11092009}
\bibinfo{author}{\bibfnamefont{N.}~\bibnamefont{Gabor}},
  \bibinfo{author}{\bibfnamefont{Z.}~\bibnamefont{Zhong}},
  \bibinfo{author}{\bibfnamefont{K.}~\bibnamefont{Bosnick}},
  \bibinfo{author}{\bibfnamefont{J.}~\bibnamefont{Park}}, \bibnamefont{and}
  \bibinfo{author}{\bibfnamefont{P.}~\bibnamefont{McEuen}},
  \bibinfo{journal}{Science} \textbf{\bibinfo{volume}{325}},
  \bibinfo{pages}{1367} (\bibinfo{year}{2009}).

\bibitem[{\citenamefont{Perebeinos and Avouris}(2006)}]{PhysRevB.74.121410}
\bibinfo{author}{\bibfnamefont{V.}~\bibnamefont{Perebeinos}} \bibnamefont{and}
  \bibinfo{author}{\bibfnamefont{P.}~\bibnamefont{Avouris}},
  \bibinfo{journal}{Phys. Rev. B} \textbf{\bibinfo{volume}{74}},
  \bibinfo{pages}{121410} (\bibinfo{year}{2006}).

\bibitem[{\citenamefont{Konabe and Okada}(2012)}]{PhysRevLett.108.227401}
\bibinfo{author}{\bibfnamefont{S.}~\bibnamefont{Konabe}} \bibnamefont{and}
  \bibinfo{author}{\bibfnamefont{S.}~\bibnamefont{Okada}},
  \bibinfo{journal}{Phys. Rev. Lett.} \textbf{\bibinfo{volume}{108}},
  \bibinfo{pages}{227401} (\bibinfo{year}{2012}).

\bibitem[{\citenamefont{Velizhanin and
  Piryatinski}(2011)}]{PhysRevLett.106.207401}
\bibinfo{author}{\bibfnamefont{K.}~\bibnamefont{Velizhanin}} \bibnamefont{and}
  \bibinfo{author}{\bibfnamefont{A.}~\bibnamefont{Piryatinski}},
  \bibinfo{journal}{Phys. Rev. Lett.} \textbf{\bibinfo{volume}{106}},
  \bibinfo{pages}{207401} (\bibinfo{year}{2011}).

\bibitem[{\citenamefont{Velizhanin and Piryatinski}(2012)}]{PhysRevB.86.165319}
\bibinfo{author}{\bibfnamefont{K.~A.} \bibnamefont{Velizhanin}}
  \bibnamefont{and}
  \bibinfo{author}{\bibfnamefont{A.}~\bibnamefont{Piryatinski}},
  \bibinfo{journal}{Phys. Rev. B} \textbf{\bibinfo{volume}{86}},
  \bibinfo{pages}{165319} (\bibinfo{year}{2012}).

\bibitem[{\citenamefont{Kilina et~al.}(2015)\citenamefont{Kilina, Kilin, and
  Tretiak}}]{doi:10.1021/acs.chemrev.5b00012}
\bibinfo{author}{\bibfnamefont{S.}~\bibnamefont{Kilina}},
  \bibinfo{author}{\bibfnamefont{D.}~\bibnamefont{Kilin}}, \bibnamefont{and}
  \bibinfo{author}{\bibfnamefont{S.}~\bibnamefont{Tretiak}},
  \bibinfo{journal}{Chemical Reviews} \textbf{\bibinfo{volume}{115}},
  \bibinfo{pages}{5929} (\bibinfo{year}{2015}).

\bibitem[{\citenamefont{Kryjevski et~al.}(2016)\citenamefont{Kryjevski,
  Gifford, Kilina, and Kilin}}]{doi:10.1063/1.4963735}
\bibinfo{author}{\bibfnamefont{A.}~\bibnamefont{Kryjevski}},
  \bibinfo{author}{\bibfnamefont{B.}~\bibnamefont{Gifford}},
  \bibinfo{author}{\bibfnamefont{S.}~\bibnamefont{Kilina}}, \bibnamefont{and}
  \bibinfo{author}{\bibfnamefont{D.}~\bibnamefont{Kilin}},
  \bibinfo{journal}{The Journal of Chemical Physics}
  \textbf{\bibinfo{volume}{145}}, \bibinfo{pages}{154112}
  (\bibinfo{year}{2016}).

\bibitem[{\citenamefont{Kryjevski et~al.}(2017)\citenamefont{Kryjevski,
  Mihaylov, Gifford, and Kilin}}]{SFJCP}
\bibinfo{author}{\bibfnamefont{A.}~\bibnamefont{Kryjevski}},
  \bibinfo{author}{\bibfnamefont{D.}~\bibnamefont{Mihaylov}},
  \bibinfo{author}{\bibfnamefont{B.}~\bibnamefont{Gifford}}, \bibnamefont{and}
  \bibinfo{author}{\bibfnamefont{D.}~\bibnamefont{Kilin}},
  \bibinfo{journal}{The Journal of Chemical Physics}
  \textbf{\bibinfo{volume}{147}}, \bibinfo{pages}{000000}
  (\bibinfo{year}{2017}).

\bibitem[{\citenamefont{Marri et~al.}(2014)\citenamefont{Marri, Govoni, and
  Ossicini}}]{doi:10.1021/ja5057328}
\bibinfo{author}{\bibfnamefont{I.}~\bibnamefont{Marri}},
  \bibinfo{author}{\bibfnamefont{M.}~\bibnamefont{Govoni}}, \bibnamefont{and}
  \bibinfo{author}{\bibfnamefont{S.}~\bibnamefont{Ossicini}},
  \bibinfo{journal}{Journal of the American Chemical Society}
  \textbf{\bibinfo{volume}{136}}, \bibinfo{pages}{13257}
  (\bibinfo{year}{2014}).

\bibitem[{\citenamefont{Kadanoff and Baym}(1962)}]{Kadanoff}
\bibinfo{author}{\bibfnamefont{L.}~\bibnamefont{Kadanoff}} \bibnamefont{and}
  \bibinfo{author}{\bibfnamefont{G.}~\bibnamefont{Baym}},
  \emph{\bibinfo{title}{Quantum Statistical Mechanics}}
  (\bibinfo{publisher}{W.A. Benjamin Inc.}, \bibinfo{address}{New York},
  \bibinfo{year}{1962}).

\bibitem[{\citenamefont{Keldysh}(1964)}]{Keldysh:1964ud}
\bibinfo{author}{\bibfnamefont{L.~V.} \bibnamefont{Keldysh}},
  \bibinfo{journal}{Zh. Eksp. Teor. Fiz.} \textbf{\bibinfo{volume}{47}},
  \bibinfo{pages}{1515} (\bibinfo{year}{1964}), \bibinfo{note}{[Sov. Phys.
  JETP20,1018(1965)]}.

\bibitem[{\citenamefont{Lifshitz and Pitaevskii}(1981)}]{Landau10}
\bibinfo{author}{\bibfnamefont{E.~M.} \bibnamefont{Lifshitz}} \bibnamefont{and}
  \bibinfo{author}{\bibfnamefont{L.~P.} \bibnamefont{Pitaevskii}},
  \emph{\bibinfo{title}{Physical Kinetics}} (\bibinfo{publisher}{Pergamon
  Press}, \bibinfo{address}{New York}, \bibinfo{year}{1981}),
  \bibinfo{edition}{1st} ed.

\bibitem[{\citenamefont{Fetter and Walecka}(1971)}]{FW}
\bibinfo{author}{\bibfnamefont{A.~L.} \bibnamefont{Fetter}} \bibnamefont{and}
  \bibinfo{author}{\bibfnamefont{J.}~\bibnamefont{Walecka}},
  \emph{\bibinfo{title}{Quantum Theory of Many-Particle Systems}}
  (\bibinfo{publisher}{McGraw-Hill}, \bibinfo{address}{New York},
  \bibinfo{year}{1971}).

\bibitem[{\citenamefont{Mahan}(1993)}]{Mahan}
\bibinfo{author}{\bibfnamefont{G.}~\bibnamefont{Mahan}},
  \emph{\bibinfo{title}{Many-Particle Physics}} (\bibinfo{publisher}{Plenum},
  \bibinfo{address}{New York, N.Y.}, \bibinfo{year}{1993}),
  \bibinfo{edition}{2nd} ed.

\bibitem[{\citenamefont{Kryjevski and Kilin}(2014)}]{molphysKK}
\bibinfo{author}{\bibfnamefont{A.}~\bibnamefont{Kryjevski}} \bibnamefont{and}
  \bibinfo{author}{\bibfnamefont{D.}~\bibnamefont{Kilin}},
  \bibinfo{journal}{Molecular Physics} \textbf{\bibinfo{volume}{112}},
  \bibinfo{pages}{430} (\bibinfo{year}{2014}).

\bibitem[{\citenamefont{Onida et~al.}(2002)\citenamefont{Onida, Reining, and
  Rubio}}]{RevModPhys.74.601}
\bibinfo{author}{\bibfnamefont{G.}~\bibnamefont{Onida}},
  \bibinfo{author}{\bibfnamefont{L.}~\bibnamefont{Reining}}, \bibnamefont{and}
  \bibinfo{author}{\bibfnamefont{A.}~\bibnamefont{Rubio}},
  \bibinfo{journal}{Rev. Mod. Phys.} \textbf{\bibinfo{volume}{74}},
  \bibinfo{pages}{601} (\bibinfo{year}{2002}).

\bibitem[{\citenamefont{K\"ummel and Kronik}(2008)}]{RevModPhys.80.3}
\bibinfo{author}{\bibfnamefont{S.}~\bibnamefont{K\"ummel}} \bibnamefont{and}
  \bibinfo{author}{\bibfnamefont{L.}~\bibnamefont{Kronik}},
  \bibinfo{journal}{Rev. Mod. Phys.} \textbf{\bibinfo{volume}{80}},
  \bibinfo{pages}{3} (\bibinfo{year}{2008}).

\bibitem[{\citenamefont{Rohlfing and Louie}(2000)}]{PhysRevB.62.4927}
\bibinfo{author}{\bibfnamefont{M.}~\bibnamefont{Rohlfing}} \bibnamefont{and}
  \bibinfo{author}{\bibfnamefont{S.}~\bibnamefont{Louie}},
  \bibinfo{journal}{Phys. Rev. B} \textbf{\bibinfo{volume}{62}},
  \bibinfo{pages}{4927} (\bibinfo{year}{2000}).

\bibitem[{\citenamefont{Strinati}(1984)}]{PhysRevB.29.5718}
\bibinfo{author}{\bibfnamefont{G.}~\bibnamefont{Strinati}},
  \bibinfo{journal}{Phys. Rev. B} \textbf{\bibinfo{volume}{29}},
  \bibinfo{pages}{5718} (\bibinfo{year}{1984}).

\bibitem[{\citenamefont{Berestetskii et~al.}(1979)\citenamefont{Berestetskii,
  Lifshitz, and Pitaevskii}}]{Berestetskii:1979aa}
\bibinfo{author}{\bibfnamefont{V.}~\bibnamefont{Berestetskii}},
  \bibinfo{author}{\bibfnamefont{E.}~\bibnamefont{Lifshitz}}, \bibnamefont{and}
  \bibinfo{author}{\bibfnamefont{L.}~\bibnamefont{Pitaevskii}},
  \emph{\bibinfo{title}{Quantum Electrodynamics}} (\bibinfo{publisher}{Oxford,
  U.K.: Pergamon Press}, \bibinfo{year}{1979}).

\bibitem[{\citenamefont{Beane et~al.}(2000)\citenamefont{Beane, Bedaque,
  Haxton, Phillips, and Savage}}]{Beane:2000fx}
\bibinfo{author}{\bibfnamefont{S.}~\bibnamefont{Beane}},
  \bibinfo{author}{\bibfnamefont{P.}~\bibnamefont{Bedaque}},
  \bibinfo{author}{\bibfnamefont{W.}~\bibnamefont{Haxton}},
  \bibinfo{author}{\bibfnamefont{D.}~\bibnamefont{Phillips}}, \bibnamefont{and}
  \bibinfo{author}{\bibfnamefont{M.}~\bibnamefont{Savage}},
  \bibinfo{journal}{Shifman, M. (ed.): At the frontier of particle physics}
  \textbf{\bibinfo{volume}{1}}, \bibinfo{pages}{133} (\bibinfo{year}{2000}).

\bibitem[{\citenamefont{Spataru et~al.}(2004)\citenamefont{Spataru,
  Ismail-Beigi, Benedict, and Louie}}]{PhysRevLett.92.077402}
\bibinfo{author}{\bibfnamefont{C.}~\bibnamefont{Spataru}},
  \bibinfo{author}{\bibfnamefont{S.}~\bibnamefont{Ismail-Beigi}},
  \bibinfo{author}{\bibfnamefont{L.}~\bibnamefont{Benedict}}, \bibnamefont{and}
  \bibinfo{author}{\bibfnamefont{S.}~\bibnamefont{Louie}},
  \bibinfo{journal}{Phys. Rev. Lett.} \textbf{\bibinfo{volume}{92}},
  \bibinfo{pages}{077402} (\bibinfo{year}{2004}).

\bibitem[{\citenamefont{Perebeinos et~al.}(2004)\citenamefont{Perebeinos,
  Tersoff, and Avouris}}]{PhysRevLett.92.257402}
\bibinfo{author}{\bibfnamefont{V.}~\bibnamefont{Perebeinos}},
  \bibinfo{author}{\bibfnamefont{J.}~\bibnamefont{Tersoff}}, \bibnamefont{and}
  \bibinfo{author}{\bibfnamefont{P.}~\bibnamefont{Avouris}},
  \bibinfo{journal}{Phys. Rev. Lett.} \textbf{\bibinfo{volume}{92}},
  \bibinfo{pages}{257402} (\bibinfo{year}{2004}).

\bibitem[{\citenamefont{Spataru et~al.}(2005)\citenamefont{Spataru,
  Ismail-Beigi, Capaz, and Louie}}]{PhysRevLett.95.247402}
\bibinfo{author}{\bibfnamefont{C.}~\bibnamefont{Spataru}},
  \bibinfo{author}{\bibfnamefont{S.}~\bibnamefont{Ismail-Beigi}},
  \bibinfo{author}{\bibfnamefont{R.}~\bibnamefont{Capaz}}, \bibnamefont{and}
  \bibinfo{author}{\bibfnamefont{S.}~\bibnamefont{Louie}},
  \bibinfo{journal}{Phys. Rev. Lett.} \textbf{\bibinfo{volume}{95}},
  \bibinfo{pages}{247402} (\bibinfo{year}{2005}).

\bibitem[{\citenamefont{\"O\ifmmode~\breve{g}\else \u{g}\fi{}\"ut
  et~al.}(2003)\citenamefont{\"O\ifmmode~\breve{g}\else \u{g}\fi{}\"ut,
  Burdick, Saad, and Chelikowsky}}]{PhysRevLett.90.127401}
\bibinfo{author}{\bibfnamefont{S.}~\bibnamefont{\"O\ifmmode~\breve{g}\else
  \u{g}\fi{}\"ut}}, \bibinfo{author}{\bibfnamefont{R.}~\bibnamefont{Burdick}},
  \bibinfo{author}{\bibfnamefont{Y.}~\bibnamefont{Saad}}, \bibnamefont{and}
  \bibinfo{author}{\bibfnamefont{J.}~\bibnamefont{Chelikowsky}},
  \bibinfo{journal}{Phys. Rev. Lett.} \textbf{\bibinfo{volume}{90}},
  \bibinfo{pages}{127401} (\bibinfo{year}{2003}).

\bibitem[{\citenamefont{Benedict et~al.}(2003)\citenamefont{Benedict, Puzder,
  Williamson, Grossman, Galli, Klepeis, Raty, and
  Pankratov}}]{PhysRevB.68.085310}
\bibinfo{author}{\bibfnamefont{L.}~\bibnamefont{Benedict}},
  \bibinfo{author}{\bibfnamefont{A.}~\bibnamefont{Puzder}},
  \bibinfo{author}{\bibfnamefont{A.}~\bibnamefont{Williamson}},
  \bibinfo{author}{\bibfnamefont{J.}~\bibnamefont{Grossman}},
  \bibinfo{author}{\bibfnamefont{G.}~\bibnamefont{Galli}},
  \bibinfo{author}{\bibfnamefont{J.}~\bibnamefont{Klepeis}},
  \bibinfo{author}{\bibfnamefont{J.-Y.} \bibnamefont{Raty}}, \bibnamefont{and}
  \bibinfo{author}{\bibfnamefont{O.}~\bibnamefont{Pankratov}},
  \bibinfo{journal}{Phys. Rev. B} \textbf{\bibinfo{volume}{68}},
  \bibinfo{pages}{085310} (\bibinfo{year}{2003}).

\bibitem[{\citenamefont{Wilson et~al.}(2009)\citenamefont{Wilson, Lu, Gygi, and
  Galli}}]{PhysRevB.79.245106}
\bibinfo{author}{\bibfnamefont{H.}~\bibnamefont{Wilson}},
  \bibinfo{author}{\bibfnamefont{D.}~\bibnamefont{Lu}},
  \bibinfo{author}{\bibfnamefont{F.}~\bibnamefont{Gygi}}, \bibnamefont{and}
  \bibinfo{author}{\bibfnamefont{G.}~\bibnamefont{Galli}},
  \bibinfo{journal}{Phys. Rev. B} \textbf{\bibinfo{volume}{79}},
  \bibinfo{pages}{245106} (\bibinfo{year}{2009}).

\bibitem[{\citenamefont{Vydrov et~al.}(2006)\citenamefont{Vydrov, Heyd, Krukau,
  and Scuseria}}]{vydrov:074106}
\bibinfo{author}{\bibfnamefont{O.}~\bibnamefont{Vydrov}},
  \bibinfo{author}{\bibfnamefont{J.}~\bibnamefont{Heyd}},
  \bibinfo{author}{\bibfnamefont{A.}~\bibnamefont{Krukau}}, \bibnamefont{and}
  \bibinfo{author}{\bibfnamefont{G.}~\bibnamefont{Scuseria}},
  \bibinfo{journal}{The Journal of Chemical Physics}
  \textbf{\bibinfo{volume}{125}}, \bibinfo{eid}{074106} (\bibinfo{year}{2006}).

\bibitem[{\citenamefont{Heyd et~al.}(2006)\citenamefont{Heyd, Scuseria, and
  Ernzerhof}}]{heyd:219906}
\bibinfo{author}{\bibfnamefont{J.}~\bibnamefont{Heyd}},
  \bibinfo{author}{\bibfnamefont{G.}~\bibnamefont{Scuseria}}, \bibnamefont{and}
  \bibinfo{author}{\bibfnamefont{M.}~\bibnamefont{Ernzerhof}},
  \bibinfo{journal}{The Journal of Chemical Physics}
  \textbf{\bibinfo{volume}{124}}, \bibinfo{eid}{219906} (\bibinfo{year}{2006}).

\bibitem[{\citenamefont{Muscat et~al.}(2001)\citenamefont{Muscat, Wander, and
  Harrison}}]{MUSCAT2001397}
\bibinfo{author}{\bibfnamefont{J.}~\bibnamefont{Muscat}},
  \bibinfo{author}{\bibfnamefont{A.}~\bibnamefont{Wander}}, \bibnamefont{and}
  \bibinfo{author}{\bibfnamefont{N.}~\bibnamefont{Harrison}},
  \bibinfo{journal}{Chemical Physics Letters} \textbf{\bibinfo{volume}{342}},
  \bibinfo{pages}{397 } (\bibinfo{year}{2001}).

\bibitem[{\citenamefont{Govoni and Galli}(2015)}]{doi:10.1021/ct500958p}
\bibinfo{author}{\bibfnamefont{M.}~\bibnamefont{Govoni}} \bibnamefont{and}
  \bibinfo{author}{\bibfnamefont{G.}~\bibnamefont{Galli}},
  \bibinfo{journal}{Journal of Chemical Theory and Computation}
  \textbf{\bibinfo{volume}{11}}, \bibinfo{pages}{2680} (\bibinfo{year}{2015}).

\bibitem[{\citenamefont{Jain et~al.}(2011)\citenamefont{Jain, Chelikowsky, and
  Louie}}]{PhysRevLett.107.216806}
\bibinfo{author}{\bibfnamefont{M.}~\bibnamefont{Jain}},
  \bibinfo{author}{\bibfnamefont{J.}~\bibnamefont{Chelikowsky}},
  \bibnamefont{and} \bibinfo{author}{\bibfnamefont{S.}~\bibnamefont{Louie}},
  \bibinfo{journal}{Phys. Rev. Lett.} \textbf{\bibinfo{volume}{107}},
  \bibinfo{pages}{216806} (\bibinfo{year}{2011}).

\bibitem[{\citenamefont{Hybertsen and Louie}(1986)}]{PhysRevB.34.5390}
\bibinfo{author}{\bibfnamefont{M.}~\bibnamefont{Hybertsen}} \bibnamefont{and}
  \bibinfo{author}{\bibfnamefont{S.}~\bibnamefont{Louie}},
  \bibinfo{journal}{Phys. Rev. B} \textbf{\bibinfo{volume}{34}},
  \bibinfo{pages}{5390} (\bibinfo{year}{1986}).

\bibitem[{\citenamefont{Noffsinger et~al.}(2012)\citenamefont{Noffsinger,
  Kioupakis, Van~de Walle, Louie, and Cohen}}]{PhysRevLett.108.167402}
\bibinfo{author}{\bibfnamefont{J.}~\bibnamefont{Noffsinger}},
  \bibinfo{author}{\bibfnamefont{E.}~\bibnamefont{Kioupakis}},
  \bibinfo{author}{\bibfnamefont{C.}~\bibnamefont{Van~de Walle}},
  \bibinfo{author}{\bibfnamefont{S.}~\bibnamefont{Louie}}, \bibnamefont{and}
  \bibinfo{author}{\bibfnamefont{M.}~\bibnamefont{Cohen}},
  \bibinfo{journal}{Phys. Rev. Lett.} \textbf{\bibinfo{volume}{108}},
  \bibinfo{pages}{167402} (\bibinfo{year}{2012}).

\bibitem[{\citenamefont{Han and Bester}(2015)}]{PhysRevB.91.085305}
\bibinfo{author}{\bibfnamefont{P.}~\bibnamefont{Han}} \bibnamefont{and}
  \bibinfo{author}{\bibfnamefont{G.}~\bibnamefont{Bester}},
  \bibinfo{journal}{Phys. Rev. B} \textbf{\bibinfo{volume}{91}},
  \bibinfo{pages}{085305} (\bibinfo{year}{2015}).

\bibitem[{\citenamefont{Kryjevski}(2015)}]{doi:10.1021/bk-2015-1196.ch010}
\bibinfo{author}{\bibfnamefont{A.}~\bibnamefont{Kryjevski}},
  \emph{\bibinfo{title}{Toward First-Principles Description of Carrier
  Relaxation in Nanoparticles in Photoinduced Processes at Surfaces and in
  Nanomaterials, ACS Symposium Series, Vol. 1196}} (\bibinfo{year}{2015}),
  chap.~\bibinfo{chapter}{10}, pp. \bibinfo{pages}{201--213}.

\bibitem[{\citenamefont{Devreese and van Camp}(1985)}]{vanCamp}
\bibinfo{author}{\bibfnamefont{D.~T.} \bibnamefont{Devreese}} \bibnamefont{and}
  \bibinfo{author}{\bibfnamefont{P.}~\bibnamefont{van Camp}},
  \emph{\bibinfo{title}{Electronic Structure, Dynamics, and Quantum Structural
  Properties of Condensed Matter}} (\bibinfo{publisher}{Springer},
  \bibinfo{address}{New York}, \bibinfo{year}{1985}).

\bibitem[{\citenamefont{Han and Bester}(2012)}]{PhysRevB.85.235422}
\bibinfo{author}{\bibfnamefont{P.}~\bibnamefont{Han}} \bibnamefont{and}
  \bibinfo{author}{\bibfnamefont{G.}~\bibnamefont{Bester}},
  \bibinfo{journal}{Phys. Rev. B} \textbf{\bibinfo{volume}{85}},
  \bibinfo{pages}{235422} (\bibinfo{year}{2012}).

\bibitem[{\citenamefont{Gonze et~al.}(2011)\citenamefont{Gonze, Boulanger, and
  Cote}}]{ANDP:ANDP201000100}
\bibinfo{author}{\bibfnamefont{X.}~\bibnamefont{Gonze}},
  \bibinfo{author}{\bibfnamefont{P.}~\bibnamefont{Boulanger}},
  \bibnamefont{and} \bibinfo{author}{\bibfnamefont{M.}~\bibnamefont{Cote}},
  \bibinfo{journal}{Annalen der Physik} \textbf{\bibinfo{volume}{523}},
  \bibinfo{pages}{168} (\bibinfo{year}{2011}).

\bibitem[{\citenamefont{Abrikosov et~al.}(1963)\citenamefont{Abrikosov, Gorkov,
  and Dzyaloshinski}}]{AGD}
\bibinfo{author}{\bibfnamefont{A.~A.} \bibnamefont{Abrikosov}},
  \bibinfo{author}{\bibfnamefont{L.}~\bibnamefont{Gorkov}}, \bibnamefont{and}
  \bibinfo{author}{\bibfnamefont{I.~E.} \bibnamefont{Dzyaloshinski}},
  \emph{\bibinfo{title}{Methods of Quantum Field Theory in Statistical
  Physics}} (\bibinfo{publisher}{Prentice-Hall}, \bibinfo{address}{Englewood
  Cliffs, NJ}, \bibinfo{year}{1963}).

\bibitem[{\citenamefont{Bl\"ochl}(1994)}]{PhysRevB.50.17953}
\bibinfo{author}{\bibfnamefont{P.~E.} \bibnamefont{Bl\"ochl}},
  \bibinfo{journal}{Phys. Rev. B} \textbf{\bibinfo{volume}{50}},
  \bibinfo{pages}{17953} (\bibinfo{year}{1994}).

\bibitem[{\citenamefont{Kresse and Joubert}(1999)}]{PhysRevB.59.1758}
\bibinfo{author}{\bibfnamefont{G.}~\bibnamefont{Kresse}} \bibnamefont{and}
  \bibinfo{author}{\bibfnamefont{D.}~\bibnamefont{Joubert}},
  \bibinfo{journal}{Phys. Rev. B} \textbf{\bibinfo{volume}{59}},
  \bibinfo{pages}{1758} (\bibinfo{year}{1999}).

\bibitem[{\citenamefont{Graham et~al.}(2012)\citenamefont{Graham, Calhoun,
  Green, Hersam, and Fleming}}]{doi:10.1021/nl2038503}
\bibinfo{author}{\bibfnamefont{M.~W.} \bibnamefont{Graham}},
  \bibinfo{author}{\bibfnamefont{T.~R.} \bibnamefont{Calhoun}},
  \bibinfo{author}{\bibfnamefont{A.~A.} \bibnamefont{Green}},
  \bibinfo{author}{\bibfnamefont{M.~C.} \bibnamefont{Hersam}},
  \bibnamefont{and} \bibinfo{author}{\bibfnamefont{G.~R.}
  \bibnamefont{Fleming}}, \bibinfo{journal}{Nano Letters}
  \textbf{\bibinfo{volume}{12}}, \bibinfo{pages}{813} (\bibinfo{year}{2012}).

\bibitem[{\citenamefont{Manzoni et~al.}(2005)\citenamefont{Manzoni, Gambetta,
  Menna, Meneghetti, Lanzani, and Cerullo}}]{PhysRevLett.94.207401}
\bibinfo{author}{\bibfnamefont{C.}~\bibnamefont{Manzoni}},
  \bibinfo{author}{\bibfnamefont{A.}~\bibnamefont{Gambetta}},
  \bibinfo{author}{\bibfnamefont{E.}~\bibnamefont{Menna}},
  \bibinfo{author}{\bibfnamefont{M.}~\bibnamefont{Meneghetti}},
  \bibinfo{author}{\bibfnamefont{G.}~\bibnamefont{Lanzani}}, \bibnamefont{and}
  \bibinfo{author}{\bibfnamefont{G.}~\bibnamefont{Cerullo}},
  \bibinfo{journal}{Phys. Rev. Lett.} \textbf{\bibinfo{volume}{94}},
  \bibinfo{pages}{207401} (\bibinfo{year}{2005}).

\bibitem[{\citenamefont{Gambetta et~al.}(2006)\citenamefont{Gambetta, Manzoni,
  Menna, Meneghetti, Cerullo, Lanzani, Tretiak, Piryatinski, Saxena, Martin
  et~al.}}]{SWCNTtau}
\bibinfo{author}{\bibfnamefont{A.}~\bibnamefont{Gambetta}},
  \bibinfo{author}{\bibfnamefont{C.}~\bibnamefont{Manzoni}},
  \bibinfo{author}{\bibfnamefont{E.}~\bibnamefont{Menna}},
  \bibinfo{author}{\bibfnamefont{M.}~\bibnamefont{Meneghetti}},
  \bibinfo{author}{\bibfnamefont{G.}~\bibnamefont{Cerullo}},
  \bibinfo{author}{\bibfnamefont{G.}~\bibnamefont{Lanzani}},
  \bibinfo{author}{\bibfnamefont{S.}~\bibnamefont{Tretiak}},
  \bibinfo{author}{\bibfnamefont{A.}~\bibnamefont{Piryatinski}},
  \bibinfo{author}{\bibfnamefont{A.}~\bibnamefont{Saxena}},
  \bibinfo{author}{\bibfnamefont{R.~L.} \bibnamefont{Martin}},
  \bibnamefont{et~al.}, \bibinfo{journal}{Nature Physics}
  \textbf{\bibinfo{volume}{2}}, \bibinfo{pages}{515} (\bibinfo{year}{2006}).

\end{thebibliography}
\end{document}